\documentclass[sigconf,nonacm]{acmart}

\usepackage{algorithmic}
\usepackage{graphicx}
\usepackage{textcomp}
\usepackage{xcolor}
\def\BibTeX{{\rm B\kern-.05em{\sc i\kern-.025em b}\kern-.08em
    T\kern-.1667em\lower.7ex\hbox{E}\kern-.125emX}}

\usepackage[frozencache,cachedir=minted-cache]{minted}
\usepackage{algorithm}
\usepackage{wrapfig}
\usepackage{comment}

\newcommand{\code}[1]{\mintinline[breaklines, breakafter=_]{C++}{#1}}

\usepackage{sc25repro}

\begin{document}

\title{Slicing Is All You Need: Towards A Universal One-Sided Algorithm for Distributed Matrix Multiplication}

\author{Benjamin Brock}
\affiliation{%
  \institution{Intel Corporation}
  \city{San Francisco}
  \state{CA}
  \country{USA}
}
\email{benjamin.brock@intel.com}

\author{Renato Golin}
\affiliation{%
  \institution{Intel Corporation}
  \city{Cambridge}
  \country{UK}
}
\email{renato.golin@intel.com}

\begin{abstract}
Many important applications across science, data analytics, and AI workloads depend on distributed matrix multiplication.  Prior work has developed a large array of algorithms suitable for different problem sizes and partitionings including 1D, 2D, 1.5D, and 2.5D algorithms.  A limitation of current work is that existing algorithms are limited to a subset of partitionings.  Multiple algorithm implementations are required to support the full space of possible partitionings.  If no algorithm implementation is available for a particular set of partitionings, one or more operands must be redistributed, increasing communication costs.  This paper presents a universal one-sided algorithm for distributed matrix multiplication that supports all combinations of partitionings and replication factors.  Our algorithm uses slicing (index arithmetic) to compute the sets of overlapping tiles that must be multiplied together.  This list of local matrix multiplies can then either be executed directly, or reordered and lowered to an optimized IR to maximize overlap.  We implement our algorithm using a high-level C++-based PGAS programming framework that performs direct GPU-to-GPU communication using intra-node interconnects.  We evaluate performance for a wide variety of partitionings and replication factors, finding that our work is competitive with PyTorch DTensor, a highly optimized distributed tensor library targeting AI models.
\end{abstract}



\maketitle

\section{Introduction}
\label{sec:intro}
Distributed matrix multiplication is an essential component of large-scale science, data analytics, and AI workloads.  There has been significant prior work on distributed matrix multiplication, which has built up a large collection of algorithms suitable for various matrix shapes, sparsities, and memory budgets.  These methods include algorithms that use various combinations of 1D and 2D partitionings, replication strategies, and sparsities.  Many previous studies have examined particular problem sizes, for example multiplying a square sparse matrix times a tall-and-skinny dense matrix~\cite{koanantakool2016}, manually implementing and comparing different variants to determine the optimal technique for a subset of problems.  Recently, distributed matrix multiplication has seen increased interest in the context of AI models, where distribution of weight matrices is often required due to limited GPU memory capacity.  Researchers in that field have determined that a 1D block outer product--style distribution with the output replicated (Megatron-LM--style tensor parallelism~\cite{megatron_lm2019}) performs particularly well for the regime of GPT-like models, while a 1D row block distribution with the second operand replicated (sequence parallelism~\cite{sequence2021}) is often necessary with large context sizes.  AI researchers have identified exploring the space of possible partitionings and replication strategies as a critical priority for AI research and have built systems that allow users to select specific partitioning and replication strategies for their weight matrices~\cite{shazeer2018meshtensorflow,xu2021gspmd,liang2022dtensor,pytorch2025dtensor}.  These systems (referred to as ``SPMD'' systems in the AI literature~\cite{xu2021gspmd}) work by allowing users to annotate distributed tensor objects with their desired distributions.  After constructing a pipeline of distributed tensor multiplications that together compose a model, the SPMD system will dispatch each individual operation to a distributed matrix multiplication algorithm, which will be selected based on its operands' dimensions and partitioning strategies.

A key limitation of current algorithms is that many different implementations are necessary to support all the desired variants of distributed matrix multiplication.  Each implementation of matrix multiplication has preconditions on its inputs: whether each operand will be 1D or 2D partitioned, whether each operand will be replicated, and how the tiles of each operand must be aligned with tiles of the other operands.  Even with a narrow set of preconditions, for example 2D block partitions with the tiles aligned and no replication for SUMMA-style algorithms~\cite{agarwal1994summa,van1997summa}, multiple implementations are necessary for different data movement strategies (Stationary A, Stationary B, or Stationary C in $C = AB$), since it is beneficial to move different matrices depending on their relative sizes (typically the largest matrix should remain stationary).  This results in a combinatorial number of implementations that are necessary to fully explore the space of possible distributed matrix multiplication techniques.

As a result, current SPMD systems dispatch to a limited set of distributed matrix multiplication implementations.  If no implementation is available for the chosen set of partitionings, as is often the case, one or more operands must be repartitioned (``resharded'') to a supported partitioning, which introduces significant communication overhead.  Due to this requirement, significant efforts have gone into developing performant and differentiable methods for resharding matrices~\cite{zhuang2024resharding,liang2025dtensorstatus}.  Still, this dependence on a fixed set of manually implemented algorithms prevents users from exploring the full space of possible partitionings.  Presenting another hurdle, different algorithms may depend on different sets of communication primitives.  This widens the set of features that vendors must implement in order to support all the various distributed matrix multiply algorithms that users' models might be dispatched to, making support more challenging.  In PyTorch's DTensor~\cite{liang2022dtensor}, the most commonly available SPMD system, some of the distributed matrix multiply algorithms depend on packed collective operations that are not available in all vendor communication backends.

Instead of dispatching to different algorithms based on distributed matrices' partitioning, an alternative strategy is to develop a single algorithm that can efficiently multiply matrices with any set of partitionings.  This paper proposes just such an algorithm: a single algorithm that can efficiently multiply matrices with \emph{any} combination of partitionings and replications, including non-aligned tiles.  Our algorithm supports the full set of 1D~\cite{fox1987matrix}, 2D~\cite{cannon1969}, 1.5D~\cite{koanantakool2016}, and 2.5D~\cite{solomonik2011} partitionings developed in the literature, as well as new, lesser-explored partitionings.  Building on previous work in one-sided algorithms for distributed matrix multiplication~\cite{krishnan2004srumma,georganas2012camatmul,brock2024}, our algorithm requires only remote get and remote accumulate communication primitives.  Our algorithm first picks a stationary matrix (A, B, or C in $C = AB$; generally the largest matrix is chosen).  Then, each process uses slicing (index arithmetic) to compute a list of local matrix multiply operations it must perform: namely, those that involve the stationary tile.  Each process then schedules this list of local operations, asynchronously retrieving any remote tiles that are required using remote get operations and asynchronously accumulating updates using remote accumulate operations.  We experiment with an exhaustive search strategy to identify the optimal schedule using a cost model, but find that a simple asynchronous execution of local operations together with an iteration offset commonly used in the literature~\cite{brock2024} is nearly optimal.  We compare our implementation with PyTorch's DTensor~\cite{liang2022dtensor}, a state-of-the-art SPMD system that supports multi-GPU distributed matrix multiplication.  We observe that our work achieves similar performance to DTensor on a range of matrix sizes derived from the MLP layers of GPT-like transformer models.  Our work supports a wide range of partitionings and replication factors with a single algorithm, achieving high performance using one-sided communication primitives across clusters of GPUs.  The core contributions of this paper are as follows:

\begin{enumerate}
  \item We present a novel one-sided algorithm for performing distributed matrix multiplication on matrices with any combinations of partitionings and replication factors.
  \item We develop a cost model that can be used to identify the best data movement strategy and schedule for a particular problem.
  \item We evaluate our technique compared to state-of-the-art methods, demonstrating that our technique performs competitively with DTensor on a range of matrix sizes derived from GPT-like models.
\end{enumerate}



\section{Background and Related Work}
\label{sec:background}
\subsection{Distributed Matrix Multiplication}
\label{sec:bkg_distributed_matmul}
Matrix multiplication computes the product of two matrices, $C = AB$.  Throughout this paper, we refer to the left operand of the distributed matrix multiply as $A$, the right operand as $B$, and the output as $C$.  Canonically, the shape of the matrix multiply is defined by the dimensions $m$, $n$, and $k$, where $A$ has shape $m \times k$, $B$ has shape $k \times n$, and $C$ has shape $m \times n$.  Distributed matrices are partitioned into multiple tiles, which are then distributed across $p$ processes.  Processes may be mapped to one or more CPU cores, or may correspond to a GPU.  Since this work is primarily GPU-focused, we use the terms process and GPU interchangeably.  Common partitions include 1D block distributions, where the matrix is partitioned across only rows or columns, and 2D distributions, where the matrix is partitioned across both dimensions.  The canonical method of partitioning a matrix, as exemplified in ScaLAPACK~\cite{blackford1997scalapack}, is to first select a tile shape, which splits the matrix into a grid of one or more tiles.  We then pick a process grid, which assigns each tile to a particular processor.  This partitioning strategy supports a wide range of matrix distributions, including 1D and 2D blocked, cyclic, and block cyclic distributions.  Traditional methods for performing distributed matrix multiplication on 2D partitioned matrices have focused on Stationary C data movement, meaning that the $A$ and $B$ matrices are communicated, while the output $C$ remains stationary.  These include Cannon's algorithm~\cite{lee1997generalized}, which performs an initial shuffle of $A$ and $B$ followed by rotating matrix tiles along rows (of $A$) and columns (of $B$), and SUMMA~\cite{agarwal1994summa,van1997summa}, which uses broadcasts within each row (of $A$) and column (of $B$) to communicate matrix tiles.  In cases where the $C$ matrix is larger or slightly smaller than $A$ and $B$, Stationary C methods are typically optimal, since they avoid accumulating remote updates to the $C$ matrix and the associated overheads.  However, when $A$ or $B$ is very large, it is usually optimal to use a Stationary A or Stationary B data movement strategy~\cite{schatz2016journey} in which the $A$ or $B$ matrix is left stationary and the other two matrices are communicated, including updates that must be accumulated to remote tiles of $C$.

While these methods reach the communication lower bound \emph{without replication}, in some cases we can further reduce communication by using replication.  Previous work has explored multiple variants of distributed matrix multiplication with one or more operands replicated; these include 2.5D algorithms, which store multiple 2D partitioned replicas~\cite{solomonik2011}, as well as 1.5D algorithms, which use 1D partitionings~\cite{koanantakool2016}.  In variants with replication, a replication factor $c$ is selected, where the replication factor must be divisible by the number of processes ($\frac{p}{c}$ is an integer).  $c$ replicas of the matrix are then created, with each replica partitioned across $\frac{p}{c}$ processes.  To perform the multiplication, a multiplication is performed within each replica using some method, with each replica performing $\frac{1}{c}$ of the total work to create a partial result.  Partial results are then reduced across replicas to produce the final result.  Note that at the extreme where $c = p$, each process has a complete copy of the matrix, while at $c = 1$, there is no replication.  The replication factor thus provides a sliding scale of replication, and prior work has found empirically that different replication factors are optimal for different problem sizes and levels of concurrency~\cite{solomonik2011,doi:10.1137/15M104253X,koanantakool2016,georganas2012}.  Replication-based communication avoiding techniques have been extended to multiple domains.  These include the problem of multi-dimensional tensors~\cite{solomonik2014massively}, the issue of selecting an optimal partitioning and replication factor for a particular problem and memory budget~\cite{cosma2019}, and sparse matrix multiplication~\cite{doi:10.1137/15M104253X,koanantakool2016}.

While most of the previously discussed work has focused on two-sided communication, particularly with the use of collectives, some prior work has investigated the use of \emph{one-sided communication} for implementing distributed matrix multiply.
One-sided communication is a model in which processes communicate with one another by directly writing to or reading from shared segments within each others' memory.  One-sided memory access is usually enabled by hardware technology, such as RDMA on an inter-node interconnect or peer access between GPUs over an intra-node network such as NVLink, Xe Link, or Infinity Fabric.
Prior work has explored one-sided 1D and 2D algorithms with replication on RDMA-equipped networks~\cite{georganas2012} as well as one-sided algorithms for sparse matrix multiplication on CPUs and GPUs~\cite{brock2024,yuxi2024}.

\subsection{Distributed Matrix Multiplication in AI}
Recent work in AI has brought renewed interest to replicated variants of distributed matrix multiplication.  While early efforts focused largely on data parallelism and pipeline parallelism~\cite{huang2019gpipe}, recent interest in very large models with weights too big to fit within a single GPU's memory have necessitated partitioning weights across multiple GPUs.  This has led to the development of several bespoke methods that target particular points in the design space of possible partitionings and replications.  In transformer models, there are multiple methods that distribute the two matrix multiplications that compose an MLP layer.  Megatron-LM--style parallelism~\cite{megatron_lm2019} performs a distributed matrix multiplication where the input data (lefthand operand $A$) is fully replicated and the first weight matrix (righthand operand $B$) is column block distributed, followed by another distributed matrix multiplication where the activation matrix (lefthand operand $A$) is column distributed and the second weight matrix (righthand operand $B$) is row distributed (an outer product--style distribution).  This has the advantage of only communicating the input data, which is typically small compared to the weight matrices.  Meanwhile, sequence parallelism~\cite{sequence2021} points out that for very large input data sizes, it may be optimal to partition the input matrix (lefthand operand $A$) into row blocks with the weights duplicated (an inner product--style distribution).  During the backward pass, the weights must be communicated.  FSDP-style parallelism~\cite{zhao2023fsdp} similarly uses an inner product--style partitioning followed by an outer product--style partitioning.  Each of these works presents a point in the larger design space of distributed matrix multiplication that works well for a particular set of matrix dimensions as well as number of processes.  However, a full exploration of the design space of 1D and 2D partitionings, replication factors, and data movement decisions remains elusive.  The algorithm presented in this paper, a method for multiplying any matrices regardless of their partitioning or replication, attempts to help tackle this problem by providing a single technique that can support a very broad range of distributions without requiring a new manual implementation of each variant.

\begin{table*}
  \caption{Primitives supported by our distributed matrix data structure.}
  \begin{tabular}{l | l}
    Definition & Description\\
    \hline
    \code{grid_shape()} & Return the shape of the matrix's tile grid.\\
    \code{tile(tile_idx, replica_idx)} & Returns view of tile \code{tile_idx} in replica \code{replica_idx}.\\
    \code{get_tile(tile_idx, replica_idx)} & Returns copy of tile \code{tile_idx} in replica \code{replica_idx}. \\
    \code{get_tile_async(tile_idx, replica_idx)} & Returns future to copy of tile. \\
    \code{accumulate_tile(replica_idx, tile_idx, view)} & Accumulate into remote tile. \\
    \code{broadcast_replica(origin_idx)} & Broadcast tiles from replica \code{origin_idx} to other replicas. \\
    \code{reduce_replicas(origin_idx)} & Accumulate values from all replicas into replica \code{origin_idx}. \\
    \code{overlapping_tiles(slice, replica_idx)} & Return list of tiles that overlap with \code{slice}. \\
    \code{tile_bounds(tile_idx)} & Return the index bounds of the tile \code{tile_idx}. \\
  \end{tabular}
  \centering
  \label{table:primitives}
\end{table*}

\section{Data Structures}
Before introducing our universal one-sided algorithm for matrix multiplication, we first describe a distributed matrix data structure as well as the primitive operations necessary for implementing matrix multiplication.  Our distributed matrix data structure is constructed using a matrix shape, a partition object that defines the partition strategy, and a replication factor.  The replication factor determines how many copies of the matrix will be created across all processes.  With $p = 12$ processes, a replication factor of 1 indicates there will be no replication, with the matrix split across all 12 processes; a replication factor of 2 indicates there will be two copies of the matrix, with each copy distributed across 6 processes; and a replication factor of 12 indicates each process will hold a copy of the entire matrix.  The partition object defines how the matrix should be distributed within each replica.  Our data structure supports high-level row block, column block, or 2D block descriptors as well as custom descriptors that define a tile shape and process grid using ScaLAPACK's conventions~\cite{blackford1997scalapack}.

Each tile of the matrix is allocated in symmetric memory so that it may be read from and written to using one-sided memory operations over an intra-node or inter-node interconnect.  Using this symmetric memory, we implement a collection of primitives operating on tiles of the matrix, as shown in Table~\ref{table:primitives}.  \code{get_tile} and \code{get_tile_async} retrieve local copies of a remote tile of the matrix, specifying the index within the tile grid of the tile to be accessed as well as the index of the replica.  The replica index is optional, and if no replica index is provided, the tile within the process' local replica is accessed.  These copy operations are host-initiated, but copy data directly between GPUs using the intra-node or inter-node interconnect.  Within a node, they will generally be executed by the runtime using the GPU's copy engine.  The \code{accumulate_tile} primitive accumulates values from a local matrix into a remote tile of the matrix held on another GPU.  When accumulating within a node over an intra-node interconnect, this is performed by launching a kernel that atomically updates each value using peer-to-peer access.  In practice, this achieves performance close to the bandwidth limit, even when using atomic operations to allow simultaneous accumulations.  When accumulating across nodes, \code{accumulate_tile} uses a coarse-grained locking scheme along with remote get and put to update the remote tile.  We also implement two methods that are necessary for synchronizing data between replicas, \code{reduce_replicas}, which reduces partially accumulated results across replicas to produce the final result, and \code{broadcast_replica}, which broadcasts tiles from an origin replica to all other replicas.  The final two primitives help perform the slicing arithmetic at the heart of our algorithm: \code{tile_bounds} and \code{overlapping_tiles}.  \code{overlapping_tiles} takes in a 2D index slice referring to an arbitrary submatrix and returns a list of tiles that overlap that slice.  \code{tile_bounds} returns the 2D slice of the matrix that a particular tile covers.

\section{Novel One-Sided Algorithm}
Our algorithm begins by selecting a data movement strategy, which is either Stationary A, Stationary B, or Stationary C.  The chosen matrix will remain stationary, while one or both of the other two matrices will be communicated.  It is usually optimal for the largest matrix to remain stationary, although the optimal choice is straightforward to verify empirically or via a cost model.  Once the data movement strategy has been chosen, each process has one or more stationary tiles, which determines what local matrix multiplication operations it must perform.

\begin{algorithm}
  \centering
\begin{minted}[linenos,fontsize=\footnotesize]{Python}
# A,B,C are distributed matrices
ops = []

# Iterate through tiles of C.
for i = 0 to C.grid_shape()[0]-1:
  for j = 0 to C.grid_shape()[1]-1:
    # If I own C.tile({i, j}), I compute its output.
    if I own C.tile({i, j}):
      c_bounds = C.tile_bounds({i, j})

      # Find all the tiles of A that must be
      # multiplied to produce my tile of C.
      a_tiles = A.overlapping_tiles({c_bounds[0],:})

      for a_idx in a_tiles:
        a_bounds = A.tile_bounds(a_idx)

        # Find all the tiles of B that need to be
        # multiplied by the A tile.
        b_tiles = B.overlapping_tiles({a_bounds[0],
                                       c_bounds[1]})

        for b_idx in b_tiles:
          b_bounds = B.tile_bounds(b_idx):

          # We now have three tiles that must be
          # multiplied; we just need to compute
          # the slices of these tiles to multiply.
          m_bound = bound(c_bounds[0], a_bounds[0])
          k_bound = bound(a_bounds[1], b_bounds[0])
          n_bound = bound(b_bounds[1], c_bounds[1])
          ops.append(op(a_idx, b_idx, {i, j},
                        {m_bound, k_bound},
                        {k_bound, n_bound},
                        {m_bound, n_bound}))
\end{minted}
  \caption{Pseudocode to generate local matrix operations with Stationary C data movement.}
  \label{alg:gen_ops}
\end{algorithm}


\subsection{Generating Local Matrix Multiply Operations}
A process must perform all the local matrix multiply operations that involve its tile of the stationary matrix, since that tile must remain in place.  This corresponds to multiplying all the tiles in the submatrices of A and B (for Stationary C), A and C (for Stationary B), or B and C (for Stationary A) that overlap with the dimensions of the stationary tile.  For example, with Stationary C data movement, if the stationary tile occupies the submatrix formed by the intersection of rows $o_m$ to $o_m+t_m$ and columns $o_n$ to $o_n+t_n$, that is $C(o_m:o_m+t_m,o_n:o_n+t_n)$, then we must multiply all tiles in the submatrix $A(o_m:o_m+t_m,0:k)$ with the corresponding tiles in the submatrix $B(0:k,o_n:o_n+t_n)$.  The algorithm used to generate the local matrix multiply operations that must be performed with Stationary C data movement is shown in Algorithm~\ref{alg:gen_ops}.  First, we identify the stationary tiles owned by a process (Line 8).  Then, for that tile, we produce a list of the tiles of A that overlap the previously mentioned submatrix $A(o_m:o_m+t_m,0:k)$ using our new \code{overlapping_tiles()} primitive (Line 13).  For each of these tiles, we examine the one or more tiles of B that it must be multiplied against (Line 20).  We then perform some index arithmetic (Lines 29-31) to compute the precise slices of these tiles that must be multiplied together\footnote{Note that there is a global-to-local offset necessary to convert the global bounds to local bounds.  This is omitted here for brevity.} (this indexing is only necessary for cases when tiles do not align).

For Stationary B data movement, shown in Algorithm~\ref{alg:gen_ops_sb}, we instead first iterate through tiles of B to find the stationary tiles, then calculate which tiles of A and C are involved in the multiplication before performing the same indexing arithmetic to compute the slices that must be multiplied.  Stationary A data movement is analogous and omitted for brevity.

A slight change is necessary when replication is used.  If the stationary matrix is replicated, each process will only compute $\frac{1}{c}$ of the work (where $c$ is the replication factor) involving its stationary tile rather than the whole set of local matrix multiplies.  As discussed in Section~\ref{sec:bkg_distributed_matmul}, this is because in algorithms with replication, each replica performs $\frac{1}{c}$ of the work to produce a partial result, with the partial results later accumulated to produce the final result.  In Algorithm~\ref{alg:gen_ops}, this entails modifying line 13 to search over $\frac{1}{c}$ of the inner dimension rather than all of it (``\code{:}'' in line 13) as well as some minor modifications to the bounds computation.

Once the list of local matrix multiplication ops has been generated, each process must then execute the local matrix multiplication operations to compute the final result.  We investigate two mechanisms: direct execution, where we largely execute the operations as-is, and lowering to an optimized IR, where we attempt to find an optimized schedule that maximizes overlap.

\begin{algorithm}
  \centering
\begin{minted}[linenos,fontsize=\footnotesize]{Python}
# A,B,C are distributed matrices
ops = []

# Iterate through tiles of B.
for k = 0 to B.grid_shape()[0]-1:
  for j = 0 to B.grid_shape()[1]-1:
    # If I own B.tile({k, j}), compute its ops.
    if I own B.tile({k, j}):
      b_bounds = B.tile_bounds({k, j})

      # Find all the tiles of A that must be
      # multiplied with my tile of B.
      a_tiles = A.overlapping_tiles({:,b_bounds[0]})

      for a_idx in a_tiles:
        a_bounds = A.tile_bounds(a_idx)

        # Find all the updates to C that are produced
        # by this multiplication.
        c_tiles = C.overlapping_tiles({a_bounds[0],
                                       b_bounds[1]})

        for c_idx in c_tiles:
          c_bounds = C.tile_bounds(c_idx):

          # We now have three tiles that must be
          # multiplied; we just need to compute
          # the slices of these tiles to multiply.
          m_bound = bound(c_bounds[0], a_bounds[0])
          k_bound = bound(a_bounds[1], b_bounds[0])
          n_bound = bound(b_bounds[1], c_bounds[1])
          ops.append(op(a_idx, b_idx, {i, j},
                        {m_bound, k_bound},
                        {k_bound, n_bound},
                        {m_bound, n_bound}))
\end{minted}
  \caption{Pseudocode to generate local matrix operations with Stationary B data movement.}
  \label{alg:gen_ops_sb}
\end{algorithm}

\begin{figure*}
  \includegraphics[width=\textwidth]{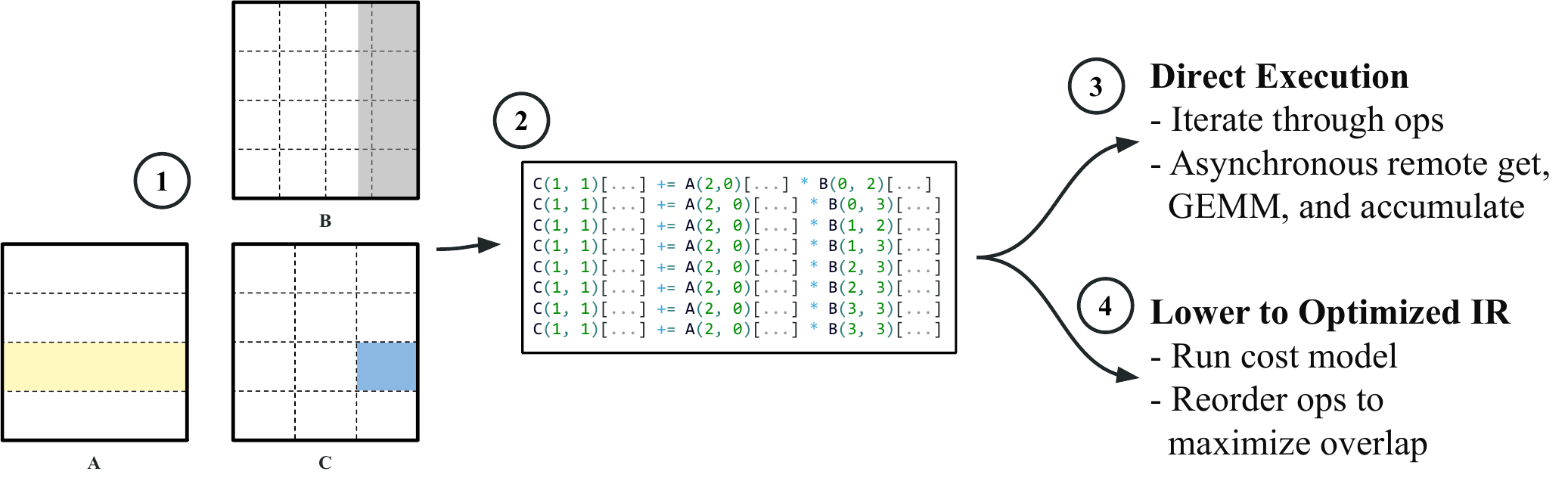}
  \caption{Our universal one-sided algorithm for distributed matrix multiplication takes in three matrices $C = AB$ with any partitioning.  This diagram shows Stationary C data movement with intentionally misaligned tiles to illustrate the algorithm's generality.  \textcircled{1} We perform slicing to identify tiles of A (yellow region) and B (grey region) that overlap with our process' stationary tile of C (blue region).  This produces a list of local matrix multiply operations \textcircled{2} that must be performed.  Note that these tiles are not required to be aligned.  We then compute the result either by \textcircled{3} directly executing this list or by \textcircled{4} reordering and lowering to an optimized IR to maximize overlap.}
  \label{fig:algorithm_diagram}
\end{figure*}

\subsection{Direct Execution}
\label{sec:direct_execution}
Once the local matrix multiplication ops have been generated, the simplest way to complete the distributed matrix multiply is to directly execute each local op in the order they are defined.  Each op references a single tile of A, B, and C, along with the slices of those tiles that must be multiplied together.  A naive process for multiplying each op is as follows:

\begin{enumerate}
  \item Obtain local copies of the tiles of A and B.  If the tile is remote, obtain a local copy with \code{get_tile()}.  If the tile is local, obtain a view of the local tile with \code{tile()}.
  \item If the C tile is remote, allocate a local matrix.  If the C tile is local, obtain a view with \code{tile()}.
  \item Perform the local matrix multiplication $C = AB$.
  \item If the tile of C is remote, issue a remote accumulation using \code{accumulate_tile()}.
\end{enumerate}

In order to achieve high performance, we augment this naive method with optimizations to improve network load balance, overlap communication and computation, allow asynchronous execution, and reduce memory allocation overhead.

First, an \emph{iteration offset} must be applied to ensure load balance.  Without an iteration offset, each process in a row or column will retrieve the same tile at the same time.  The iteration offset changes the order in which each process executes to balance network communication.  For our iteration offset, we simply add the two indices of the stationary tile as done in previous work~\cite{brock2024}.

Second, we implement \emph{prefetching} to allow overlap between communication and computation.  Before issuing a local matrix multiplication operation, we prefetch the next two tiles that will be needed using \code{get_tile_async()}, which asynchronously fetches a tile over the network, returning a future object.  In the next iteration, we will synchronize on each future object, returning the tile objects.


Third, we \emph{launch all work asynchronously} to allow for additional overlap.  In the case that remote updates must be applied to tiles of C, we asynchronously launch a local GEMM, then a kernel that accumulates the result into the remote tile, with the accumulation kernel dependent on the local GEMM.  We can then continue without synchronizing on either the GEMM or accumulate kernel, allowing GEMMs and accumulates from different iterations to execute in parallel.  We allow a configurable number of concurrent GEMMs and concurrent accumulates, respectively, with higher limits permitting more asynchrony at the cost of potentially higher memory usage.

Fourth, we use a \emph{memory pool} to avoid the overheads associated with GPU memory allocation, which can be significant and often induce synchronization across the whole device.  Our memory pool performs one initial GPU memory allocation, then allocates memory from the CPU side.


Note that \emph{no code changes are necessary} to support replication of any combination of A, B, and C.  This algorithm transparently benefits from replication, since each process accesses its local replica by default in calls to \code{get_tile()} and \code{accumulate_tile()}.  This allows this algorithm to support all combinations of replication factors for A, B, and C, including unusual combinations not supported by traditional algorithms.  In the case that C is replicated, a call to \code{reduce_replicas()} after the algorithm completes reduces the partial results computed in each replica to produce the final result.

As discussed in Section~\ref{sec:eval}, we find that direct execution achieves a high percentage of theoretical peak FLOPs and is competitive with high performance libraries.

\subsection{Lowering to Optimized IR}
Instead of directly executing the ops generated by our algorithm, we can instead lower these ops into another intermediate representation (IR) in which communication is explicit.  This lowering process will determine the optimal order for executing these tiles, as well as how to overlap communication and computation.  The lowering is a two step process.  First, we build a computation graph for each process representing the local component matrix multiplications it must perform as well as the matrix tiles these component operations are dependent upon.  The computation graph is a bipartite graph with compute operations on one side and data on the other.  Each component operation has edges to the tiles it depends upon, representing a data dependency that may require communication.  Data dependency edges have labels representing whether the dependency is satisfied.  If a matrix tile happens to be local to the process, the edge is created in a satisfied state.  Otherwise, the edge will start out in an unsatisfied state and will be changed to satisfied as the computation graph is traversed during IR generation.

To generate an IR with explicitly overlapped communication and computation, we traverse the computation graph to identify available work.  The output IR ops consist of a list of zero or more compute operations and zero or more communication operations, with the total amount of concurrent compute and communication configurable via a hyperparameter.  If all the data dependency edges associated with a particular component operation are in the satisfied state, that compute operation can be scheduled in the current output IR op being generated.  Once any eligible compute is scheduled, we can then add communication operations corresponding to data dependencies in the unsatisfied state.  Once the output IR op is processed, the data dependency edges corresponding to any communication performed will be set to the satisfied state so that their dependent computations can be scheduled in the next output IR op. We use several techniques to generate IR, both a greedy technique and techniques based on a cost model.

\subsubsection*{Greedy Algorithm}
In the greedy algorithm, any eligible compute is first scheduled.  Then, any eligible communication is scheduled.  Both compute and communication are scheduled in each output op up to the maximum limits set as hyperparameters.

\subsubsection*{Cost Model--Based Methods}
We can also use a cost model to attempt to pick an optimal lowering.  Given an output IR op, which consists of overlapped communication and computation operations, we can estimate the runtime cost of that op as the maximum of the communication and computation cost.  The computation cost we estimate using a simple Roofline model based on the matrix tile size as well as our GPU's arithmetic peak and memory bandwidth peak.  Communication cost we can estimate by taking the number of bytes that must be fetched in each communication operation and dividing it by the bandwidth available between the process and remote tile.  Note that remote tiles on different GPUs may have different bandwidth available, depending on network topology.

We implement two different output IR generation strategies using our cost model.  The first is a greedy approach, which fully fills each output IR op to the available communication and computation limit, selecting which communication and computation to pick based on the cost model.  The second approach is an exhaustive search, which examines every possible selection of output ops, using the cost model to estimate the total cost, picking the output with the lowest cost.

\section{Evaluation}
\label{sec:eval}

In order to evaluate the performance of our universal one-sided distributed matrix multiplication algorithm, we implemented it using a fork of Distributed Ranges~\cite{brock2024dr}, a C++ library for distributed GPU programming.  We then performed a series of experiments to compare our algorithm, using a variety of partitionings and replication factors, to PyTorch DTensor~\cite{pytorch2025dtensor}, a production SPMD system.

\subsection{Implementation}

We implement our universal one-sided algorithm using a fork of Distributed Ranges~\cite{brock2024dr} that supports distributed GPU communication using Intel SHMEM~\cite{brooks2024ishmem} and \mbox{NVSHMEM}~\cite{langer2022nvshmem}.  Our framework is able to compile using SYCL and CUDA, depending on the system, and uses the vendor-provided oneMKL and cuBLAS libraries to perform local matrix multiplies.  Within a single node, we perform accumulations using a hand-written kernel, which uses atomic operations to accumulate to a remote tile.  We observed that our accumulate kernel achieves about 80\% of the bandwidth achieved by the copy engine, which is in line with expectations.  Both backends support asynchronous execution.  The SYCL backend uses a SYCL out-of-order queue to launch work, explicitly creating dependencies on SYCL events to enforce ordering between operations when necessary, namely to force ordering between a local GEMM that produces a result and the accumulate kernel that accumulates it as described in Section~\ref{sec:direct_execution}.  A similar effect is achieved in CUDA by using CUDA events to enforce synchronization between events on separate CUDA streams.

\subsection{Experiments}

\begin{table}
  \caption{System details for the PVC and H100 systems used in our experiments, detailing number of devices, per-device unidirectional link bandwidth available via the intra-node interconnect (Xe~Link or NVLink), and per-device theoretical 32-bit floating point peak.}
\resizebox{\linewidth}{!}{%
  \begin{tabular}{l l r r r}
  \toprule
  System & Number of Devices & Link BW & FP32 Peak\\
  \midrule
  PVC & 12 & 26.5 GB/s & 22.7 TFLOPs\\
  H100 & 8 & 450 GB/s & 67 TFLOPs\\

  \bottomrule
  \end{tabular}
}
  \label{table:system_config}
\end{table}

\begin{figure*}
  \includegraphics[width=\columnwidth]{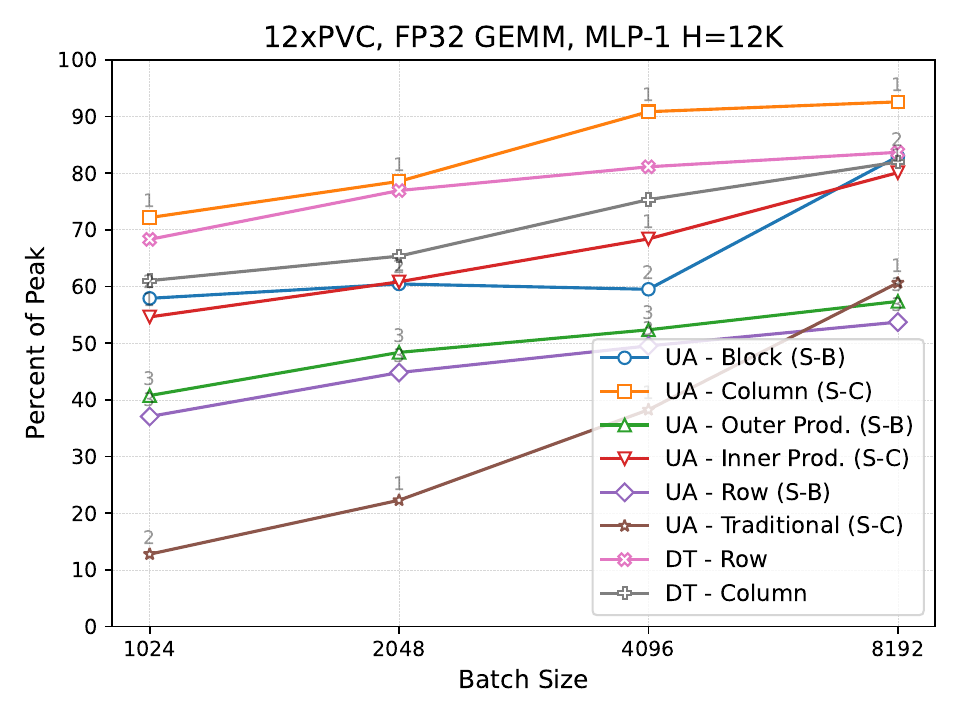}
  \includegraphics[width=\columnwidth]{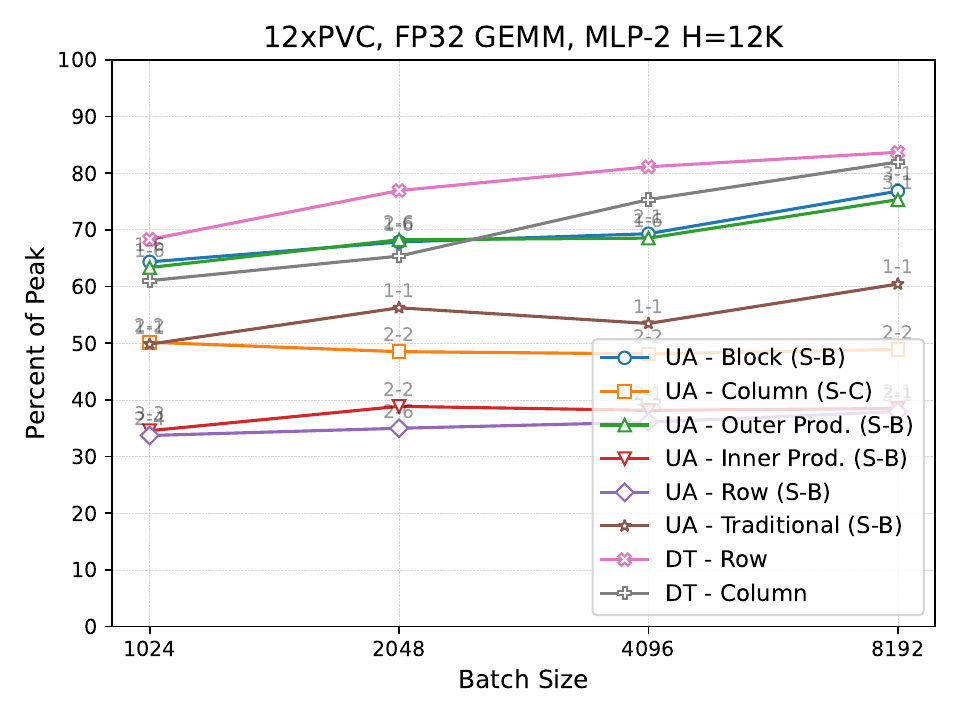}
  \caption{Experiments on an Intel PVC system comparing our methods versus DTensor for matrix multiplications with dimensions reflective of the MLP layer in a GPT-like transformer.  (MLP-1: $m = \textnormal{batch size}$, $n = 48\textnormal{K}$, $k = 12\textnormal{K}$, MLP-2: $m = \textnormal{batch size}$, $n = 12\textnormal{K}$, $k = 48\textnormal{K}$.)  Replication factors are plotted above each result.  For MLP-2, where we used mixed replication factors, the replication factor for $A$ and $B$ is shown before the dash and replication factor for $C$ is shown after the dash.  S-C refers to Stationary C data movement, while S-B refers to Stationary B data movement.}
  \label{fig:mlp1}
  \label{fig:mlp2}
\end{figure*}

We ran our experiments on two systems: a system equipped with Intel GPUs (``PVC'') and one equipped with Nvidia GPUs (``H100'').  Selected details for both systems are shown in Table~\ref{table:system_config}.  The Intel PVC system is equipped with 6 Intel Data Center GPU Max 1550 GPUs, codename Ponte Vecchio (PVC).  Each PVC GPU has two tiles, each of which we use as an independent device.  Each PVC tile has 64 GB of HBM2e memory.  The tiles on each GPU can access each others' memory using an inter-tile interconnect with 230 GB/s of theoretical unidirectional bandwidth.  All twelve tiles within a node can also access each others' memory using an intra-node Xe Link interconnect with 20 GB/s of unidirectional link bandwidth.  Each tile has a 32-bit single precision floating point theoretical peak of 22.7 TFLOPs.  Our system is equipped with two Intel Xeon CPU Max 9470 CPUs and 1 TB of memory.  All C++ code was compiled with the Intel oneAPI C++ compiler version 2024.2.0 as well as Intel SHMEM 1.5.0.  For the PyTorch DTensor experiments, we used PyTorch nightly version 2.9.0.dev20250714+xpu with Intel GPU support and the xccl backend, which uses the Intel oneCCL library for GPU communication.

The Nvidia H100 system is equipped with 8 Nvidia H100 GPUs.  Each H100 GPU has 80 GB of HBM3 memory.  The H100 GPUs are connected by an intra-node NVLink interconnect with 450 GB/s of unidirectional link bandwidth.  Each GPU has a a 32-bit single precision floating point theoretical peak of 67 TFLOPs.  Our system is equipped with two Intel Xeon Platinum 8480C CPUs and 2TB of memory.  All C++ code was compiled with \code{nvcc} using CUDA Toolkit version 12.8 and NVSHMEM 3.2.5.  For the PyTorch Dtensor experiments, we used PyTorch 2.7.1 and the nccl backend.

Our method supports any combinations of partitionings and replication factors for A, B, and C, which presents a combinatorially large set of parameters to test.  For our algorithm, we exhaustively test all combinations of row block, column block, and rectangular 2D block with all valid replication factors.  We also test a ``traditional'' 2D blocked partitioning in which tiles of A, B, and C must be aligned, a common restriction of traditional distributed matrix multiplication methods.  Due to issues of space, here we only highlight a subset of partitionings of interest.  For all benchmarks, we report the best of 10 consecutive runs.

In all of the experiments presented here, we use direct execution for our algorithm.  In our early experiments, we observed that generating an optimal schedule had significant impact for problems with misaligned tiles, which produce iterations with variable amounts of compute and communication.  This creates more opportunities for rearranging work to provide better overlap, since a naive execution might overlap a small amount of compute with a large amount of communication or vice versa, sacrificing opportunities for overlap.  However, after implementing the optimizations described in Section~\ref{sec:direct_execution}, particularly allowing asynchrony between iterations, direct execution was almost always as efficient as the optimal schedule.

For DTensor, we report row and column partitionings, which were the fastest partitionings in our testing.  We observed faster performance without replication, so numbers without replication are shown here.  It should be noted that DTensor did not support multiplying many partitionings due to either no available distributed matrix multiply implementation (we observed this for operands with different replication factors) or requiring packed collectives that are not available from all vendors (we observed this with 2D partitionings).  We benchmarked DTensor by calling \code{torch.matmul()} on two DTensor objects.  If any dimension of the output tensor is left as a series of partial products that must be reduced, we issue a \code{redistribute()} command to complete the reduction and produce the final result, converting the DTensor placement of the unreduced dimension from \code{Partial} to \code{Shard}.

For an additional baseline, on the Nvidia H100 system we also compare our performance to COSMA~\cite{cosma2019}.  Given a particular problem size, number of processors, and memory capacity, COSMA automatically selects a matrix partitioning and replication factor to minimize communication within a memory budget.  COSMA automatically scales between 2D (no replication) and 2.5D algorithms (with replication).  Here, we run COSMA with its NCCL backend.  We achieved higher performance with computation/communication overlap turned off, so we report those numbers here.  We allow COSMA an unlimited memory budget.

\subsubsection{GPT-Like Problem}

In order to evaluate our one-sided matrix multiplication algorithm on a range of real-world problem sizes, we conducted a series of distributed matrix multiplications using sizes derived from the multi-layer perceptron (MLP) layer of GPT-like models.  The MLP layer takes in data with a hidden dimension size $h$, then applies two linear layers (matrix multiplications), the first multiplication by a short and fat matrix of dimension $h \times rh$, increasing the hidden dimension to $rh$ ($r$ is most commonly 4), followed by the second multiplication by a tall and skinny matrix of dimension $rh \times h$, which decreases the hidden dimension back to its original size.  We refer to these two problem sizes as MLP-1 and MLP-2 and choose $r = 4$.  Note that for these experiments, we consider only partitionings that do not entirely eliminate communication.  All experiments are run across all 12 PVC tiles.

We first turn to the MLP-1 layer, which is a matrix multiplication with $m = \text{batch size}, n = 48K, k = 12K$.  This first multiplication in the MLP takes a batch of data with some hidden dimension (here 12K) and expands it (here to 48K).  For each partitioning strategy, we report the replication factor that achieved the highest performance as well as the data movement strategy that achieved the highest performance on the largest batch size.

Looking first at the performance results on the Intel PVC system shown in the left plot of Figure~\ref{fig:mlp1}, we observe that the highest performing partitionings for our algorithm (``UA'') are column block and inner product--style (row block times column block) partitioning.  These algorithm achieve the highest performance because they only require moving the A matrix, which is the smallest matrix for MLP-1.  Note that although they are both fastest with the Stationary C code path, both the C and B matrices are stationary for these partitionings---the advantage of 1D partitionings over 2D partitionings for irregular problem sizes is that they only require moving one matrix, while 2D partitionings require moving two matrices.  While column block and inner product both have the same communication volume, column block performs better because its local GEMMs are more efficient.  The inner product partitioning ends up multiplying thin row and column panels together to produce a small square output block, while the column block partitioning multiplies a column panel by a square block to produce a column panel.  While the 2D block distribution improves in performance as the matrix multiplication becomes larger and more square, it is held back by the fact that it must move both the small A matrix and larger C matrix.  Note that the optimal replication factor is greater than 1 for some of the poorly performing partitionings, which must move the large B matrix.  This is because they see benefit from reducing their larger communication volume. The column block distribution, which only moves the small A matrix, is able to overlap almost all of its communication, thus seeing no benefit from replication, which introduces additional accumulation overhead in the required call to \code{reduce_replicas()}.

While determining the precise algorithm that DTensor uses is difficult due to its complex dispatching logic, it appears to strongly prefer to perform outer product--style matrix multiplies where C is accumulated across processes, even if one of the operands must be repartitioned.  This is likely due to the fact that matrix multiplies must be dispatched to a limited number of algorithms.  With a partitioning that minimizes data movement, our algorithm is competitive with DTensor, even outperforming it for our best-performing partitioning.

\begin{figure*}
  \includegraphics[width=\columnwidth]{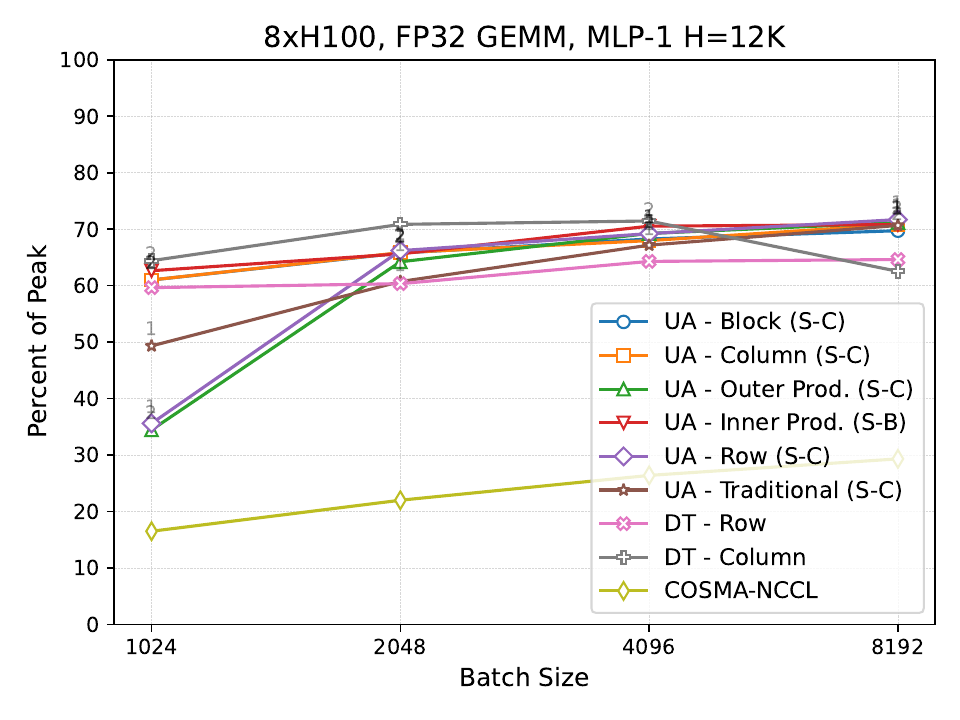}
  \includegraphics[width=\columnwidth]{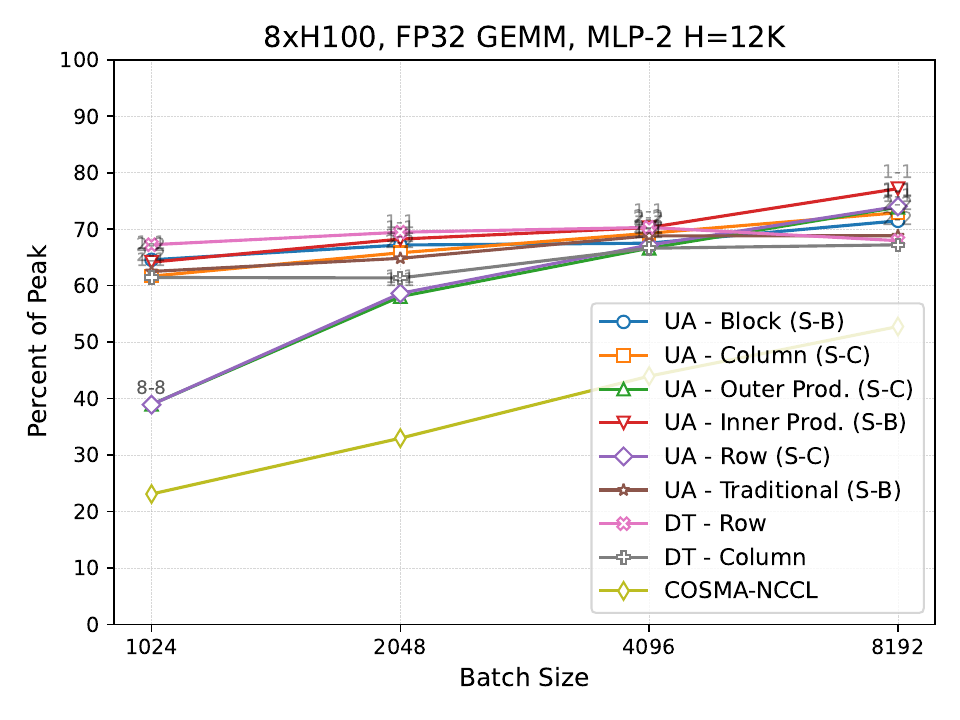}
  \caption{Experiments on an Nvidia H100 system comparing our methods versus DTensor for matrix multiplications with dimensions reflective of the MLP layer in a GPT-like transformer.  (MLP-1: $m = \textnormal{batch size}$, $n = 48\textnormal{K}$, $k = 12\textnormal{K}$, MLP-2: $m = \textnormal{batch size}$, $n = 12\textnormal{K}$, $k = 48\textnormal{K}$.)  Replication factors are plotted above each result.  For MLP-2, where we used mixed replication factors, the replication factor for $A$ and $B$ is shown before the dash and replication factor for $C$ is shown after the dash.  S-C refers to Stationary C data movement, while S-B refers to Stationary B data movement.}
  \label{fig:h100_mlp1}
  \label{fig:h100_mlp2}
\end{figure*}

Looking at the results for MLP-1 on the Nvidia H100 system, shown in the left plot of Figure~\ref{fig:h100_mlp1}, we see that there is a much smaller performance difference between different partitionings for our method (the lines are closer together than in Figure~\ref{fig:mlp1}).  This is likely because of the higher link bandwidth available on the H100 system, as shown in Table~\ref{table:system_config}, which means that communication is less of a bottleneck.  Still, the column and inner product partitionings perform best, particularly for lower batch sizes.  Our algorithm (``UA'') is competitive with DTensor across a range of partitionings.

COSMA performs poorly on MLP-1.  It picks a 2D distribution with replication that only requires a final group AllReduce to reduce replicas of the C matrix.  While this has a relatively low communication volume, it is possible the performance of this group collective is suboptimal.

We now turn to MLP-2, where $m = \text{batch size}, n = 12K, k = 48K$.  This second multiplication in the MLP takes in a batch of the expanded hidden dimension (here 48K) and shrinks it back down to its original size (here 12K).

We begin with the Intel PVC results in the right plot in Figure~\ref{fig:mlp2}. Here, the MLP reduces the hidden dimension back to its original size, meaning the output C matrix is the smallest.  This means outer product--style algorithms, which communicate by accumulating the output C matrix, will perform better.  For our algorithm, we observe that an outer product--style distribution (column block times row block) as well as a 2D block distribution perform the best.  Both of these distributions avoid moving the large B weight matrix.  Note that we see better performance with higher replication factors here.  This is likely for two reasons: 1) remote accumulation has somewhat lower performance than remote get, meaning that there is greater benefit from reducing communication volume, and 2) replication increases the size of the tiles within each replica, leading to greater arithmetic intensity and higher local GEMM performance.  With an outer product partitioning, we noticed a tradeoff: without replication, local GEMM performance was low due to suboptimal local GEMM sizes; with a high replication factor, local GEMM performance was very high, but performance was impacted by high accumulation overhead.  The optimal replication factor shown in the plot is a happy medium between those two extremes.  We observed that using a different replication factor for the output matrix compared to the input matrices achieved better performance, likely due to lower accumulation overhead and the opportunity to overlap the accumulations that are required, since a low replication factor in the output results in multiple small accumulations.  Our performance does not quite match DTensor's, coming within 5\%.  Replacing our custom accumulation kernel with an optimized oneCCL collective, which is what DTensor is utilizing, would likely breach the gap.  Further exploring and optimizing local GEMM performance might also increase performance.

Looking at the results for MLP-2 on the Nvidia H100 system, shown in the right plot in Figure~\ref{fig:h100_mlp2}, we again notice a smaller difference in performance between different partitionings.  Interestingly, we see that the outer product partitioning, which was the fastest technique on the PVC system, performs poorly on H100, with Stationary C data movement, which here moves the larger A matrix rather than C matrix, winning out.  Analyzing performance data, it appears that our accumulation kernel is impacting the performance of the local GEMMs on H100, which was not an issue we observed on PVC.  Previous work has observed the propensity for communication kernels to interfere with computation~\cite{changho2023ark,agrawal2025c3,deepseek2025v3}, using persistent kernels that occupy a limited number of SMs as well as using DMA copy engines instead of kernels for communication.  It's likely that optimizing our accumulate kernel using these techniques would result in higher performance for Stationary B data movement.  Still, many partitionings achieved higher performance despite performing more communication, with the best performing method generally matching or exceeding DTensor's performance.

\section{Conclusion and Future Work}
This paper presents a one-sided algorithm for multiplying dense matrices with any mix of partitionings and replication factors.  This helps enable exploration of the full design space of possible distributions.  In the future, we plan to integrate our algorithm into a production SPMD system such as DTensor in order to further expand the number of supported distributions.  This work does not address the issue of how to select an optimal partitioning for a particular problem.  Prior work~\cite{cosma2019} has developed techniques for automatically selecting an appropriate distribution for a particular problem size, and it should be possible to combine the work presented here with similar techniques.

\bibliographystyle{ACM-Reference-Format}
\bibliography{references}


\begin{thebibliography}{33}


\ifx \showCODEN    \undefined \def \showCODEN     #1{\unskip}     \fi
\ifx \showDOI      \undefined \def \showDOI       #1{#1}\fi
\ifx \showISBNx    \undefined \def \showISBNx     #1{\unskip}     \fi
\ifx \showISBNxiii \undefined \def \showISBNxiii  #1{\unskip}     \fi
\ifx \showISSN     \undefined \def \showISSN      #1{\unskip}     \fi
\ifx \showLCCN     \undefined \def \showLCCN      #1{\unskip}     \fi
\ifx \shownote     \undefined \def \shownote      #1{#1}          \fi
\ifx \showarticletitle \undefined \def \showarticletitle #1{#1}   \fi
\ifx \showURL      \undefined \def \showURL       {\relax}        \fi
\providecommand\bibfield[2]{#2}
\providecommand\bibinfo[2]{#2}
\providecommand\natexlab[1]{#1}
\providecommand\showeprint[2][]{arXiv:#2}

\bibitem[Agarwal et~al\mbox{.}(1994)]%
        {agarwal1994summa}
\bibfield{author}{\bibinfo{person}{R.~C. Agarwal}, \bibinfo{person}{F.~G.
  Gustavson}, {and} \bibinfo{person}{M. Zubair}.}
  \bibinfo{year}{1994}\natexlab{}.
\newblock \showarticletitle{A high-performance matrix-multiplication algorithm
  on a distributed-memory parallel computer, using overlapped communication}.
\newblock \bibinfo{journal}{\emph{{IBM} Journal of Research and Development}}
  \bibinfo{volume}{38}, \bibinfo{number}{6} (\bibinfo{year}{1994}),
  \bibinfo{pages}{673--681}.
\newblock
\urldef\tempurl%
\url{https://doi.org/10.1147/rd.386.0673}
\showDOI{\tempurl}


\bibitem[Agrawal et~al\mbox{.}(2025)]%
        {agrawal2025c3}
\bibfield{author}{\bibinfo{person}{Anirudha Agrawal}, \bibinfo{person}{Shaizeen
  Aga}, \bibinfo{person}{Suchita Pati}, {and} \bibinfo{person}{Mahzabeen
  Islam}.} \bibinfo{year}{2025}\natexlab{}.
\newblock \bibinfo{title}{Optimizing {ML} Concurrent Computation and
  Communication with {GPU} {DMA} Engines}.
\newblock
\newblock
\showeprint[arxiv]{2412.14335}~[cs.AR]
\urldef\tempurl%
\url{https://arxiv.org/abs/2412.14335}
\showURL{%
\tempurl}


\bibitem[Azad et~al\mbox{.}(2016)]%
        {doi:10.1137/15M104253X}
\bibfield{author}{\bibinfo{person}{Ariful Azad}, \bibinfo{person}{Grey
  Ballard}, \bibinfo{person}{Aydin Bulu\c{c}}, \bibinfo{person}{James Demmel},
  \bibinfo{person}{Laura Grigori}, \bibinfo{person}{Oded Schwartz},
  \bibinfo{person}{Sivan Toledo}, {and} \bibinfo{person}{Samuel Williams}.}
  \bibinfo{year}{2016}\natexlab{}.
\newblock \showarticletitle{Exploiting Multiple Levels of Parallelism in Sparse
  Matrix-Matrix Multiplication}.
\newblock \bibinfo{journal}{\emph{SIAM Journal on Scientific Computing}}
  \bibinfo{volume}{38}, \bibinfo{number}{6} (\bibinfo{year}{2016}),
  \bibinfo{pages}{C624--C651}.
\newblock
\urldef\tempurl%
\url{https://doi.org/10.1137/15M104253X}
\showDOI{\tempurl}
\showeprint{https://doi.org/10.1137/15M104253X}


\bibitem[Blackford et~al\mbox{.}(1997)]%
        {blackford1997scalapack}
\bibfield{author}{\bibinfo{person}{L~Susan Blackford},
  \bibinfo{person}{Jaeyoung Choi}, \bibinfo{person}{Andy Cleary},
  \bibinfo{person}{Eduardo D'Azevedo}, \bibinfo{person}{James Demmel},
  \bibinfo{person}{Inderjit Dhillon}, \bibinfo{person}{Jack Dongarra},
  \bibinfo{person}{Sven Hammarling}, \bibinfo{person}{Greg Henry},
  \bibinfo{person}{Antoine Petitet}, {et~al\mbox{.}}}
  \bibinfo{year}{1997}\natexlab{}.
\newblock \bibinfo{booktitle}{\emph{{ScaLAPACK} Users' Guide}}.
\newblock \bibinfo{publisher}{SIAM}.
\newblock


\bibitem[Brock et~al\mbox{.}(2024a)]%
        {brock2024}
\bibfield{author}{\bibinfo{person}{Benjamin Brock}, \bibinfo{person}{Ayd\i{}n
  Bulu\c{c}}, {and} \bibinfo{person}{Katherine Yelick}.}
  \bibinfo{year}{2024}\natexlab{a}.
\newblock \showarticletitle{{RDMA}-Based Algorithms for Sparse Matrix
  Multiplication on {GPU}s}. In \bibinfo{booktitle}{\emph{Proceedings of the
  38th ACM International Conference on Supercomputing}} (Kyoto, Japan)
  \emph{(\bibinfo{series}{ICS '24})}. \bibinfo{publisher}{Association for
  Computing Machinery}, \bibinfo{address}{New York, NY, USA},
  \bibinfo{pages}{225–235}.
\newblock
\showISBNx{9798400706103}
\urldef\tempurl%
\url{https://doi.org/10.1145/3650200.3656623}
\showDOI{\tempurl}


\bibitem[Brock et~al\mbox{.}(2024b)]%
        {brock2024dr}
\bibfield{author}{\bibinfo{person}{Benjamin Brock}, \bibinfo{person}{Robert
  Cohn}, \bibinfo{person}{Suyash Bakshi}, \bibinfo{person}{Tuomas Karna},
  \bibinfo{person}{Jeongnim Kim}, \bibinfo{person}{Mateusz Nowak},
  \bibinfo{person}{\L{}ukasz undefinedlusarczyk}, \bibinfo{person}{Kacper
  Stefanski}, {and} \bibinfo{person}{Timothy~G. Mattson}.}
  \bibinfo{year}{2024}\natexlab{b}.
\newblock \showarticletitle{{Distributed Ranges}: A Model for Distributed Data
  Structures, Algorithms, and Views}. In \bibinfo{booktitle}{\emph{Proceedings
  of the 38th ACM International Conference on Supercomputing}} (Kyoto, Japan)
  \emph{(\bibinfo{series}{ICS '24})}. \bibinfo{publisher}{Association for
  Computing Machinery}, \bibinfo{address}{New York, NY, USA},
  \bibinfo{pages}{236–246}.
\newblock
\showISBNx{9798400706103}
\urldef\tempurl%
\url{https://doi.org/10.1145/3650200.3656632}
\showDOI{\tempurl}


\bibitem[Brooks et~al\mbox{.}(2024)]%
        {brooks2024ishmem}
\bibfield{author}{\bibinfo{person}{Alex Brooks}, \bibinfo{person}{Philip
  Marshall}, \bibinfo{person}{David Ozog}, \bibinfo{person}{Md.
  Wasi-Ur-Rahman}, \bibinfo{person}{Lawrence Stewart}, {and}
  \bibinfo{person}{Rithwik Tom}.} \bibinfo{year}{2024}\natexlab{}.
\newblock \showarticletitle{{Intel}\textsuperscript{®} {SHMEM}:
  {GPU}-initiated {OpenSHMEM} using {SYCL}}. In
  \bibinfo{booktitle}{\emph{SC24-W: Workshops of the International Conference
  for High Performance Computing, Networking, Storage and Analysis}}.
  \bibinfo{pages}{1288--1301}.
\newblock
\urldef\tempurl%
\url{https://doi.org/10.1109/SCW63240.2024.00169}
\showDOI{\tempurl}


\bibitem[Cannon(1969)]%
        {cannon1969}
\bibfield{author}{\bibinfo{person}{Lynn~Elliot Cannon}.}
  \bibinfo{year}{1969}\natexlab{}.
\newblock \emph{\bibinfo{title}{A cellular computer to implement the {Kalman}
  filter algorithm}}.
\newblock \bibinfo{thesistype}{Ph.\,D. Dissertation}. \bibinfo{school}{Montana
  State University}, \bibinfo{address}{USA}.
\newblock
\newblock
\shownote{AAI7010025}.


\bibitem[{DeepSeek-AI} et~al\mbox{.}(2025)]%
        {deepseek2025v3}
\bibfield{author}{\bibinfo{person}{{DeepSeek-AI}}, \bibinfo{person}{Aixin Liu},
  \bibinfo{person}{Bei Feng}, \bibinfo{person}{Bing Xue},
  \bibinfo{person}{Bingxuan Wang}, \bibinfo{person}{Bochao Wu},
  \bibinfo{person}{Chengda Lu}, \bibinfo{person}{Chenggang Zhao},
  \bibinfo{person}{Chengqi Deng}, \bibinfo{person}{Chenyu Zhang},
  \bibinfo{person}{Chong Ruan}, \bibinfo{person}{Damai Dai},
  \bibinfo{person}{Daya Guo}, \bibinfo{person}{Dejian Yang},
  \bibinfo{person}{Deli Chen}, \bibinfo{person}{Dongjie Ji},
  \bibinfo{person}{Erhang Li}, \bibinfo{person}{Fangyun Lin},
  \bibinfo{person}{Fucong Dai}, \bibinfo{person}{Fuli Luo},
  \bibinfo{person}{Guangbo Hao}, \bibinfo{person}{Guanting Chen},
  \bibinfo{person}{Guowei Li}, \bibinfo{person}{H. Zhang}, \bibinfo{person}{Han
  Bao}, \bibinfo{person}{Hanwei Xu}, \bibinfo{person}{Haocheng Wang},
  \bibinfo{person}{Haowei Zhang}, \bibinfo{person}{Honghui Ding},
  \bibinfo{person}{Huajian Xin}, \bibinfo{person}{Huazuo Gao},
  \bibinfo{person}{Hui Li}, \bibinfo{person}{Hui Qu}, \bibinfo{person}{J.~L.
  Cai}, \bibinfo{person}{Jian Liang}, \bibinfo{person}{Jianzhong Guo},
  \bibinfo{person}{Jiaqi Ni}, \bibinfo{person}{Jiashi Li},
  \bibinfo{person}{Jiawei Wang}, \bibinfo{person}{Jin Chen},
  \bibinfo{person}{Jingchang Chen}, \bibinfo{person}{Jingyang Yuan},
  \bibinfo{person}{Junjie Qiu}, \bibinfo{person}{Junlong Li},
  \bibinfo{person}{Junxiao Song}, \bibinfo{person}{Kai Dong},
  \bibinfo{person}{Kai Hu}, \bibinfo{person}{Kaige Gao}, \bibinfo{person}{Kang
  Guan}, \bibinfo{person}{Kexin Huang}, \bibinfo{person}{Kuai Yu},
  \bibinfo{person}{Lean Wang}, \bibinfo{person}{Lecong Zhang},
  \bibinfo{person}{Lei Xu}, \bibinfo{person}{Leyi Xia}, \bibinfo{person}{Liang
  Zhao}, \bibinfo{person}{Litong Wang}, \bibinfo{person}{Liyue Zhang},
  \bibinfo{person}{Meng Li}, \bibinfo{person}{Miaojun Wang},
  \bibinfo{person}{Mingchuan Zhang}, \bibinfo{person}{Minghua Zhang},
  \bibinfo{person}{Minghui Tang}, \bibinfo{person}{Mingming Li},
  \bibinfo{person}{Ning Tian}, \bibinfo{person}{Panpan Huang},
  \bibinfo{person}{Peiyi Wang}, \bibinfo{person}{Peng Zhang},
  \bibinfo{person}{Qiancheng Wang}, \bibinfo{person}{Qihao Zhu},
  \bibinfo{person}{Qinyu Chen}, \bibinfo{person}{Qiushi Du},
  \bibinfo{person}{R.~J. Chen}, \bibinfo{person}{R.~L. Jin},
  \bibinfo{person}{Ruiqi Ge}, \bibinfo{person}{Ruisong Zhang},
  \bibinfo{person}{Ruizhe Pan}, \bibinfo{person}{Runji Wang},
  \bibinfo{person}{Runxin Xu}, \bibinfo{person}{Ruoyu Zhang},
  \bibinfo{person}{Ruyi Chen}, \bibinfo{person}{S.~S. Li},
  \bibinfo{person}{Shanghao Lu}, \bibinfo{person}{Shangyan Zhou},
  \bibinfo{person}{Shanhuang Chen}, \bibinfo{person}{Shaoqing Wu},
  \bibinfo{person}{Shengfeng Ye}, \bibinfo{person}{Shengfeng Ye},
  \bibinfo{person}{Shirong Ma}, \bibinfo{person}{Shiyu Wang},
  \bibinfo{person}{Shuang Zhou}, \bibinfo{person}{Shuiping Yu},
  \bibinfo{person}{Shunfeng Zhou}, \bibinfo{person}{Shuting Pan},
  \bibinfo{person}{T. Wang}, \bibinfo{person}{Tao Yun}, \bibinfo{person}{Tian
  Pei}, \bibinfo{person}{Tianyu Sun}, \bibinfo{person}{W.~L. Xiao},
  \bibinfo{person}{Wangding Zeng}, \bibinfo{person}{Wanjia Zhao},
  \bibinfo{person}{Wei An}, \bibinfo{person}{Wen Liu}, \bibinfo{person}{Wenfeng
  Liang}, \bibinfo{person}{Wenjun Gao}, \bibinfo{person}{Wenqin Yu},
  \bibinfo{person}{Wentao Zhang}, \bibinfo{person}{X.~Q. Li},
  \bibinfo{person}{Xiangyue Jin}, \bibinfo{person}{Xianzu Wang},
  \bibinfo{person}{Xiao Bi}, \bibinfo{person}{Xiaodong Liu},
  \bibinfo{person}{Xiaohan Wang}, \bibinfo{person}{Xiaojin Shen},
  \bibinfo{person}{Xiaokang Chen}, \bibinfo{person}{Xiaokang Zhang},
  \bibinfo{person}{Xiaosha Chen}, \bibinfo{person}{Xiaotao Nie},
  \bibinfo{person}{Xiaowen Sun}, \bibinfo{person}{Xiaoxiang Wang},
  \bibinfo{person}{Xin Cheng}, \bibinfo{person}{Xin Liu}, \bibinfo{person}{Xin
  Xie}, \bibinfo{person}{Xingchao Liu}, \bibinfo{person}{Xingkai Yu},
  \bibinfo{person}{Xinnan Song}, \bibinfo{person}{Xinxia Shan},
  \bibinfo{person}{Xinyi Zhou}, \bibinfo{person}{Xinyu Yang},
  \bibinfo{person}{Xinyuan Li}, \bibinfo{person}{Xuecheng Su},
  \bibinfo{person}{Xuheng Lin}, \bibinfo{person}{Y.~K. Li},
  \bibinfo{person}{Y.~Q. Wang}, \bibinfo{person}{Y.~X. Wei},
  \bibinfo{person}{Y.~X. Zhu}, \bibinfo{person}{Yang Zhang},
  \bibinfo{person}{Yanhong Xu}, \bibinfo{person}{Yanhong Xu},
  \bibinfo{person}{Yanping Huang}, \bibinfo{person}{Yao Li},
  \bibinfo{person}{Yao Zhao}, \bibinfo{person}{Yaofeng Sun},
  \bibinfo{person}{Yaohui Li}, \bibinfo{person}{Yaohui Wang},
  \bibinfo{person}{Yi Yu}, \bibinfo{person}{Yi Zheng}, \bibinfo{person}{Yichao
  Zhang}, \bibinfo{person}{Yifan Shi}, \bibinfo{person}{Yiliang Xiong},
  \bibinfo{person}{Ying He}, \bibinfo{person}{Ying Tang},
  \bibinfo{person}{Yishi Piao}, \bibinfo{person}{Yisong Wang},
  \bibinfo{person}{Yixuan Tan}, \bibinfo{person}{Yiyang Ma},
  \bibinfo{person}{Yiyuan Liu}, \bibinfo{person}{Yongqiang Guo},
  \bibinfo{person}{Yu Wu}, \bibinfo{person}{Yuan Ou}, \bibinfo{person}{Yuchen
  Zhu}, \bibinfo{person}{Yuduan Wang}, \bibinfo{person}{Yue Gong},
  \bibinfo{person}{Yuheng Zou}, \bibinfo{person}{Yujia He},
  \bibinfo{person}{Yukun Zha}, \bibinfo{person}{Yunfan Xiong},
  \bibinfo{person}{Yunxian Ma}, \bibinfo{person}{Yuting Yan},
  \bibinfo{person}{Yuxiang Luo}, \bibinfo{person}{Yuxiang You},
  \bibinfo{person}{Yuxuan Liu}, \bibinfo{person}{Yuyang Zhou},
  \bibinfo{person}{Z.~F. Wu}, \bibinfo{person}{Z.~Z. Ren},
  \bibinfo{person}{Zehui Ren}, \bibinfo{person}{Zhangli Sha},
  \bibinfo{person}{Zhe Fu}, \bibinfo{person}{Zhean Xu}, \bibinfo{person}{Zhen
  Huang}, \bibinfo{person}{Zhen Zhang}, \bibinfo{person}{Zhenda Xie},
  \bibinfo{person}{Zhengyan Zhang}, \bibinfo{person}{Zhewen Hao},
  \bibinfo{person}{Zhibin Gou}, \bibinfo{person}{Zhicheng Ma},
  \bibinfo{person}{Zhigang Yan}, \bibinfo{person}{Zhihong Shao},
  \bibinfo{person}{Zhipeng Xu}, \bibinfo{person}{Zhiyu Wu},
  \bibinfo{person}{Zhongyu Zhang}, \bibinfo{person}{Zhuoshu Li},
  \bibinfo{person}{Zihui Gu}, \bibinfo{person}{Zijia Zhu},
  \bibinfo{person}{Zijun Liu}, \bibinfo{person}{Zilin Li},
  \bibinfo{person}{Ziwei Xie}, \bibinfo{person}{Ziyang Song},
  \bibinfo{person}{Ziyi Gao}, {and} \bibinfo{person}{Zizheng Pan}.}
  \bibinfo{year}{2025}\natexlab{}.
\newblock \bibinfo{title}{{DeepSeek-V3} Technical Report}.
\newblock
\newblock
\showeprint[arxiv]{2412.19437}~[cs.CL]
\urldef\tempurl%
\url{https://arxiv.org/abs/2412.19437}
\showURL{%
\tempurl}


\bibitem[Fox et~al\mbox{.}(1987)]%
        {fox1987matrix}
\bibfield{author}{\bibinfo{person}{Geoffrey~C. Fox}, \bibinfo{person}{Steve~W.
  Otto}, {and} \bibinfo{person}{Anthony J.~G. Hey}.}
  \bibinfo{year}{1987}\natexlab{}.
\newblock \showarticletitle{Matrix Algorithms on a Hypercube I: Matrix
  Multiplication}.
\newblock \bibinfo{journal}{\emph{Parallel Comput.}} \bibinfo{volume}{4},
  \bibinfo{number}{1} (\bibinfo{year}{1987}), \bibinfo{pages}{17--31}.
\newblock
\urldef\tempurl%
\url{https://doi.org/10.1016/0167-8191(87)80009-5}
\showDOI{\tempurl}


\bibitem[Georganas et~al\mbox{.}(2012a)]%
        {georganas2012}
\bibfield{author}{\bibinfo{person}{Evangelos Georganas}, \bibinfo{person}{Jorge
  Gonz\'{a}lez-Dom\'{\i}nguez}, \bibinfo{person}{Edgar Solomonik},
  \bibinfo{person}{Yili Zheng}, \bibinfo{person}{Juan Touri\~{n}o}, {and}
  \bibinfo{person}{Katherine Yelick}.} \bibinfo{year}{2012}\natexlab{a}.
\newblock \showarticletitle{Communication Avoiding and Overlapping for
  Numerical Linear Algebra}. In \bibinfo{booktitle}{\emph{Proceedings of the
  International Conference on High Performance Computing, Networking, Storage
  and Analysis}} (Salt Lake City, Utah) \emph{(\bibinfo{series}{SC '12})}.
  \bibinfo{publisher}{IEEE Computer Society Press},
  \bibinfo{address}{Washington, DC, USA}, Article \bibinfo{articleno}{100},
  \bibinfo{numpages}{11}~pages.
\newblock
\showISBNx{9781467308045}


\bibitem[Georganas et~al\mbox{.}(2012b)]%
        {georganas2012camatmul}
\bibfield{author}{\bibinfo{person}{Evangelos Georganas}, \bibinfo{person}{Jorge
  Gonzalez-Dominguez}, \bibinfo{person}{Edgar Solomonik}, \bibinfo{person}{Yili
  Zheng}, \bibinfo{person}{Juan Tourino}, {and} \bibinfo{person}{Katherine
  Yelick}.} \bibinfo{year}{2012}\natexlab{b}.
\newblock \showarticletitle{Communication avoiding and overlapping for
  numerical linear algebra}. In \bibinfo{booktitle}{\emph{{SC}'12: Proceedings
  of the International Conference on High Performance Computing, Networking,
  Storage and Analysis}}. \bibinfo{pages}{1--11}.
\newblock
\urldef\tempurl%
\url{https://doi.org/10.1109/SC.2012.32}
\showDOI{\tempurl}


\bibitem[Hong and Bulu\c{c}(2024)]%
        {yuxi2024}
\bibfield{author}{\bibinfo{person}{Yuxi Hong} {and} \bibinfo{person}{Aydin
  Bulu\c{c}}.} \bibinfo{year}{2024}\natexlab{}.
\newblock \showarticletitle{A Sparsity-Aware Distributed-Memory Algorithm for
  Sparse-Sparse Matrix Multiplication}. In
  \bibinfo{booktitle}{\emph{Proceedings of the International Conference for
  High Performance Computing, Networking, Storage, and Analysis}} (Atlanta, GA,
  USA) \emph{(\bibinfo{series}{SC '24})}. \bibinfo{publisher}{IEEE Press},
  Article \bibinfo{articleno}{47}, \bibinfo{numpages}{14}~pages.
\newblock
\showISBNx{9798350352917}
\urldef\tempurl%
\url{https://doi.org/10.1109/SC41406.2024.00053}
\showDOI{\tempurl}


\bibitem[Huang et~al\mbox{.}(2019)]%
        {huang2019gpipe}
\bibfield{author}{\bibinfo{person}{Yanping Huang}, \bibinfo{person}{Youlong
  Cheng}, \bibinfo{person}{Ankur Bapna}, \bibinfo{person}{Orhan Firat},
  \bibinfo{person}{Mia~Xu Chen}, \bibinfo{person}{Dehao Chen},
  \bibinfo{person}{HyoukJoong Lee}, \bibinfo{person}{Jiquan Ngiam},
  \bibinfo{person}{Quoc~V. Le}, \bibinfo{person}{Yonghui Wu}, {and}
  \bibinfo{person}{Zhifeng Chen}.} \bibinfo{year}{2019}\natexlab{}.
\newblock \bibinfo{booktitle}{\emph{{GPipe}: efficient training of giant neural
  networks using pipeline parallelism}}.
\newblock \bibinfo{publisher}{Curran Associates Inc.}, \bibinfo{address}{Red
  Hook, NY, USA}.
\newblock


\bibitem[Hwang et~al\mbox{.}(2023)]%
        {changho2023ark}
\bibfield{author}{\bibinfo{person}{Changho Hwang}, \bibinfo{person}{KyoungSoo
  Park}, \bibinfo{person}{Ran Shu}, \bibinfo{person}{Xinyuan Qu},
  \bibinfo{person}{Peng Cheng}, {and} \bibinfo{person}{Yongqiang Xiong}.}
  \bibinfo{year}{2023}\natexlab{}.
\newblock \showarticletitle{{ARK}: {GPU-driven} Code Execution for Distributed
  Deep Learning}. In \bibinfo{booktitle}{\emph{20th USENIX Symposium on
  Networked Systems Design and Implementation (NSDI 23)}}.
  \bibinfo{publisher}{USENIX Association}, \bibinfo{address}{Boston, MA},
  \bibinfo{pages}{87--101}.
\newblock
\showISBNx{978-1-939133-33-5}
\urldef\tempurl%
\url{https://www.usenix.org/conference/nsdi23/presentation/hwang}
\showURL{%
\tempurl}


\bibitem[Koanantakool et~al\mbox{.}(2016)]%
        {koanantakool2016}
\bibfield{author}{\bibinfo{person}{Penporn Koanantakool},
  \bibinfo{person}{Ariful Azad}, \bibinfo{person}{Aydin Buluç},
  \bibinfo{person}{Dmitriy Morozov}, \bibinfo{person}{Sang-Yun Oh},
  \bibinfo{person}{Leonid Oliker}, {and} \bibinfo{person}{Katherine Yelick}.}
  \bibinfo{year}{2016}\natexlab{}.
\newblock \showarticletitle{Communication-Avoiding Parallel Sparse-Dense
  Matrix-Matrix Multiplication}. In \bibinfo{booktitle}{\emph{2016 IEEE
  International Parallel and Distributed Processing Symposium (IPDPS)}}.
  \bibinfo{pages}{842--853}.
\newblock
\urldef\tempurl%
\url{https://doi.org/10.1109/IPDPS.2016.117}
\showDOI{\tempurl}


\bibitem[Krishnan and Nieplocha(2004)]%
        {krishnan2004srumma}
\bibfield{author}{\bibinfo{person}{Manojkumar Krishnan} {and}
  \bibinfo{person}{J. Nieplocha}.} \bibinfo{year}{2004}\natexlab{}.
\newblock \showarticletitle{{SRUMMA}: A Matrix Multiplication Algorithm
  Suitable for Clusters and Scalable Shared Memory Systems}. In
  \bibinfo{booktitle}{\emph{18th International Parallel and Distributed
  Processing Symposium, 2004. Proceedings.}} \bibinfo{pages}{70--}.
\newblock
\urldef\tempurl%
\url{https://doi.org/10.1109/IPDPS.2004.1303000}
\showDOI{\tempurl}


\bibitem[Kwasniewski et~al\mbox{.}(2019)]%
        {cosma2019}
\bibfield{author}{\bibinfo{person}{Grzegorz Kwasniewski},
  \bibinfo{person}{Marko Kabi\'{c}}, \bibinfo{person}{Maciej Besta},
  \bibinfo{person}{Joost VandeVondele}, \bibinfo{person}{Raffaele Solc\`{a}},
  {and} \bibinfo{person}{Torsten Hoefler}.} \bibinfo{year}{2019}\natexlab{}.
\newblock \showarticletitle{Red-Blue Pebbling Revisited: Near Optimal Parallel
  Matrix-Matrix Multiplication}. In \bibinfo{booktitle}{\emph{Proceedings of
  the International Conference for High Performance Computing, Networking,
  Storage and Analysis}} (Denver, Colorado) \emph{(\bibinfo{series}{SC '19})}.
  \bibinfo{publisher}{Association for Computing Machinery},
  \bibinfo{address}{New York, NY, USA}, Article \bibinfo{articleno}{24},
  \bibinfo{numpages}{22}~pages.
\newblock
\showISBNx{9781450362290}
\urldef\tempurl%
\url{https://doi.org/10.1145/3295500.3356181}
\showDOI{\tempurl}


\bibitem[Langer et~al\mbox{.}(2022)]%
        {langer2022nvshmem}
\bibfield{author}{\bibinfo{person}{Akhil Langer}, \bibinfo{person}{Seth
  Howell}, \bibinfo{person}{Sreeram Potluri}, \bibinfo{person}{Jim Dinan},
  {and} \bibinfo{person}{Jiri Kraus}.} \bibinfo{year}{2022}\natexlab{}.
\newblock \showarticletitle{Dynamic Symmetric Heap Allocation in {NVSHMEM}}. In
  \bibinfo{booktitle}{\emph{{OpenSHMEM} and Related Technologies. {OpenSHMEM}
  in the Era of Exascale and Smart Networks}},
  \bibfield{editor}{\bibinfo{person}{Stephen Poole}, \bibinfo{person}{Oscar
  Hernandez}, \bibinfo{person}{Matthew Baker}, {and} \bibinfo{person}{Tony
  Curtis}} (Eds.). \bibinfo{publisher}{Springer International Publishing},
  \bibinfo{address}{Cham}, \bibinfo{pages}{187--198}.
\newblock
\showISBNx{978-3-031-04888-3}


\bibitem[Lee et~al\mbox{.}(1997)]%
        {lee1997generalized}
\bibfield{author}{\bibinfo{person}{Hyuk-Jae Lee}, \bibinfo{person}{James~P
  Robertson}, {and} \bibinfo{person}{Jos{\'e}~AB Fortes}.}
  \bibinfo{year}{1997}\natexlab{}.
\newblock \showarticletitle{Generalized {Cannon}'s algorithm for parallel
  matrix multiplication}. In \bibinfo{booktitle}{\emph{Proceedings of the 11th
  international conference on Supercomputing}}. \bibinfo{pages}{44--51}.
\newblock


\bibitem[Li et~al\mbox{.}(2021)]%
        {sequence2021}
\bibfield{author}{\bibinfo{person}{Shenggui Li}, \bibinfo{person}{Fuzhao Xue},
  \bibinfo{person}{Yongbin Li}, {and} \bibinfo{person}{Yang You}.}
  \bibinfo{year}{2021}\natexlab{}.
\newblock \showarticletitle{Sequence Parallelism: Making {4D} Parallelism
  Possible}.
\newblock \bibinfo{journal}{\emph{CoRR}}  \bibinfo{volume}{abs/2105.13120}
  (\bibinfo{year}{2021}).
\newblock
\showeprint[arXiv]{2105.13120}
\urldef\tempurl%
\url{https://arxiv.org/abs/2105.13120}
\showURL{%
\tempurl}


\bibitem[Liang(2022)]%
        {liang2022dtensor}
\bibfield{author}{\bibinfo{person}{Wanchao Liang}.}
  \bibinfo{year}{2022}\natexlab{}.
\newblock \bibinfo{title}{{RFC}: {PyTorch} {DistributedTensor}}.
\newblock
  \bibinfo{howpublished}{\url{https://dev-discuss.pytorch.org/t/rfc-pytorch-distributedtensor/740}}.
\newblock
\newblock
\shownote{{PyTorch} Developer Mailing List, accessed 28 July 2025}.


\bibitem[Liang(2025)]%
        {liang2025dtensorstatus}
\bibfield{author}{\bibinfo{person}{Wanchao Liang}.}
  \bibinfo{year}{2025}\natexlab{}.
\newblock \bibinfo{title}{{DTensor} — Status, Design and Looking Forward}.
\newblock
  \bibinfo{howpublished}{\url{https://dev-discuss.pytorch.org/t/dtensor-status-design-and-looking-forward/2749}}.
\newblock
\newblock
\shownote{{PyTorch} Developer Mailing List, accessed 29 July 2025}.


\bibitem[{PyTorch} Core Team(2025)]%
        {pytorch2025dtensor}
{PyTorch} Core Team \bibinfo{year}{2025}\natexlab{}.
\newblock \bibinfo{booktitle}{\emph{\texttt{torch.distributed.tensor} —
  {PyTorch} 2.7 Documentation}}.
\newblock {PyTorch} Core Team.
\newblock
\newblock
\shownote{{PyTorch} 2.7 stable documentation, accessed 28 July 2025}.


\bibitem[Schatz et~al\mbox{.}(2016)]%
        {schatz2016journey}
\bibfield{author}{\bibinfo{person}{Martin~D. Schatz},
  \bibinfo{person}{Robert~A. van~de Geijn}, {and} \bibinfo{person}{Jack
  Poulson}.} \bibinfo{year}{2016}\natexlab{}.
\newblock \showarticletitle{Parallel Matrix Multiplication: A Systematic
  Journey}.
\newblock \bibinfo{journal}{\emph{SIAM Journal on Scientific Computing}}
  \bibinfo{volume}{38}, \bibinfo{number}{6} (\bibinfo{year}{2016}),
  \bibinfo{pages}{C748--C781}.
\newblock
\urldef\tempurl%
\url{https://doi.org/10.1137/140993478}
\showDOI{\tempurl}
\showeprint{https://doi.org/10.1137/140993478}


\bibitem[Shazeer et~al\mbox{.}(2018)]%
        {shazeer2018meshtensorflow}
\bibfield{author}{\bibinfo{person}{Noam Shazeer}, \bibinfo{person}{Youlong
  Cheng}, \bibinfo{person}{Niki Parmar}, \bibinfo{person}{Dustin Tran},
  \bibinfo{person}{Ashish Vaswani}, \bibinfo{person}{Penporn Koanantakool},
  \bibinfo{person}{Peter Hawkins}, \bibinfo{person}{HyoukJoong Lee},
  \bibinfo{person}{Mingsheng Hong}, \bibinfo{person}{Cliff Young},
  \bibinfo{person}{Ryan Sepassi}, {and} \bibinfo{person}{Blake Hechtman}.}
  \bibinfo{year}{2018}\natexlab{}.
\newblock \bibinfo{title}{{Mesh-TensorFlow}: Deep Learning for Supercomputers}.
\newblock
\newblock
\showeprint[arxiv]{1811.02084}~[cs.LG]
\urldef\tempurl%
\url{https://arxiv.org/abs/1811.02084}
\showURL{%
\tempurl}


\bibitem[Shoeybi et~al\mbox{.}(2019)]%
        {megatron_lm2019}
\bibfield{author}{\bibinfo{person}{Mohammad Shoeybi}, \bibinfo{person}{Mostofa
  Patwary}, \bibinfo{person}{Raul Puri}, \bibinfo{person}{Patrick LeGresley},
  \bibinfo{person}{Jared Casper}, {and} \bibinfo{person}{Bryan Catanzaro}.}
  \bibinfo{year}{2019}\natexlab{}.
\newblock \showarticletitle{{Megatron-LM}: Training Multi-Billion Parameter
  Language Models Using Model Parallelism}.
\newblock \bibinfo{journal}{\emph{CoRR}}  \bibinfo{volume}{abs/1909.08053}
  (\bibinfo{year}{2019}).
\newblock
\showeprint[arXiv]{1909.08053}
\urldef\tempurl%
\url{http://arxiv.org/abs/1909.08053}
\showURL{%
\tempurl}


\bibitem[Solomonik and Demmel(2011)]%
        {solomonik2011}
\bibfield{author}{\bibinfo{person}{Edgar Solomonik} {and}
  \bibinfo{person}{James Demmel}.} \bibinfo{year}{2011}\natexlab{}.
\newblock \showarticletitle{Communication-Optimal Parallel {2.5D} Matrix
  Multiplication and {LU} Factorization Algorithms}. In
  \bibinfo{booktitle}{\emph{Euro-Par 2011 Parallel Processing}},
  \bibfield{editor}{\bibinfo{person}{Emmanuel Jeannot},
  \bibinfo{person}{Raymond Namyst}, {and} \bibinfo{person}{Jean Roman}} (Eds.).
  \bibinfo{publisher}{Springer Berlin Heidelberg}, \bibinfo{address}{Berlin,
  Heidelberg}, \bibinfo{pages}{90--109}.
\newblock
\showISBNx{978-3-642-23397-5}


\bibitem[Solomonik et~al\mbox{.}(2014)]%
        {solomonik2014massively}
\bibfield{author}{\bibinfo{person}{Edgar Solomonik}, \bibinfo{person}{Devin
  Matthews}, \bibinfo{person}{Jeff~R Hammond}, \bibinfo{person}{John~F
  Stanton}, {and} \bibinfo{person}{James Demmel}.}
  \bibinfo{year}{2014}\natexlab{}.
\newblock \showarticletitle{A Massively Parallel Tensor Contraction Framework
  for Coupled-Cluster Computations}.
\newblock \bibinfo{journal}{\emph{J. Parallel and Distrib. Comput.}}
  \bibinfo{volume}{74}, \bibinfo{number}{12} (\bibinfo{year}{2014}),
  \bibinfo{pages}{3176--3190}.
\newblock


\bibitem[Van De~Geijn and Watts(1997)]%
        {van1997summa}
\bibfield{author}{\bibinfo{person}{Robert~A Van De~Geijn} {and}
  \bibinfo{person}{Jerrell Watts}.} \bibinfo{year}{1997}\natexlab{}.
\newblock \showarticletitle{{SUMMA}: Scalable universal matrix multiplication
  algorithm}.
\newblock \bibinfo{journal}{\emph{Concurrency: Practice and Experience}}
  \bibinfo{volume}{9}, \bibinfo{number}{4} (\bibinfo{year}{1997}),
  \bibinfo{pages}{255--274}.
\newblock


\bibitem[Xu et~al\mbox{.}(2021)]%
        {xu2021gspmd}
\bibfield{author}{\bibinfo{person}{Yuanzhong Xu}, \bibinfo{person}{HyoukJoong
  Lee}, \bibinfo{person}{Dehao Chen}, \bibinfo{person}{Blake Hechtman},
  \bibinfo{person}{Yanping Huang}, \bibinfo{person}{Rahul Joshi},
  \bibinfo{person}{Maxim Krikun}, \bibinfo{person}{Dmitry Lepikhin},
  \bibinfo{person}{Andy Ly}, \bibinfo{person}{Marcello Maggioni},
  \bibinfo{person}{Ruoming Pang}, \bibinfo{person}{Noam Shazeer},
  \bibinfo{person}{Shibo Wang}, \bibinfo{person}{Tao Wang},
  \bibinfo{person}{Yonghui Wu}, {and} \bibinfo{person}{Zhifeng Chen}.}
  \bibinfo{year}{2021}\natexlab{}.
\newblock \bibinfo{title}{{GSPMD}: General and Scalable Parallelization for
  {ML} Computation Graphs}.
\newblock
\newblock
\showeprint[arxiv]{2105.04663}~[cs.DC]
\urldef\tempurl%
\url{https://arxiv.org/abs/2105.04663}
\showURL{%
\tempurl}


\bibitem[Zhao et~al\mbox{.}(2023)]%
        {zhao2023fsdp}
\bibfield{author}{\bibinfo{person}{Yanli Zhao}, \bibinfo{person}{Andrew Gu},
  \bibinfo{person}{Rohan Varma}, \bibinfo{person}{Liang Luo},
  \bibinfo{person}{Chien-Chin Huang}, \bibinfo{person}{Min Xu},
  \bibinfo{person}{Less Wright}, \bibinfo{person}{Hamid Shojanazeri},
  \bibinfo{person}{Myle Ott}, \bibinfo{person}{Sam Shleifer},
  \bibinfo{person}{Alban Desmaison}, \bibinfo{person}{Can Balioglu},
  \bibinfo{person}{Pritam Damania}, \bibinfo{person}{Bernard Nguyen},
  \bibinfo{person}{Geeta Chauhan}, \bibinfo{person}{Yuchen Hao},
  \bibinfo{person}{Ajit Mathews}, {and} \bibinfo{person}{Shen Li}.}
  \bibinfo{year}{2023}\natexlab{}.
\newblock \bibinfo{title}{{PyTorch} {FSDP}: Experiences on Scaling Fully
  Sharded Data Parallel}.
\newblock
\newblock
\showeprint[arxiv]{2304.11277}~[cs.DC]
\urldef\tempurl%
\url{https://arxiv.org/abs/2304.11277}
\showURL{%
\tempurl}


\bibitem[Zhuang et~al\mbox{.}(2024)]%
        {zhuang2024resharding}
\bibfield{author}{\bibinfo{person}{Yonghao Zhuang}, \bibinfo{person}{Hexu
  Zhao}, \bibinfo{person}{Lianmin Zheng}, \bibinfo{person}{Zhuohan Li},
  \bibinfo{person}{Eric~P. Xing}, \bibinfo{person}{Qirong Ho},
  \bibinfo{person}{Joseph~E. Gonzalez}, \bibinfo{person}{Ion Stoica}, {and}
  \bibinfo{person}{Hao Zhang}.} \bibinfo{year}{2024}\natexlab{}.
\newblock \bibinfo{title}{On Optimizing the Communication of Model
  Parallelism}.
\newblock
\newblock
\showeprint[arxiv]{2211.05322}~[cs.LG]
\urldef\tempurl%
\url{https://arxiv.org/abs/2211.05322}
\showURL{%
\tempurl}


\end{thebibliography}

\scriptsize
\noindent
\newline Optimization Notice: Software and workloads used in
performance tests may have been optimized for performance only on
Intel microprocessors.  Performance tests, such as SYSmark and
MobileMark, are measured using specific computer systems,
components, software, operations and functions.  Any change to any
of those factors may cause the results to vary.  You should
consult other information and performance tests to assist you in
fully evaluating your contemplated purchases, including the
performance of that product when combined with other products.
For more information go to \url{http://www.intel.com/performance}.

\noindent Intel, Xeon, and Intel Xeon Phi are trademarks of Intel Corporation in the U.S. and/or other countries.

\normalsize


\clearpage
\appendix

\twocolumn[%
{\begin{center}
\Huge
Appendix: Artifact Description/Artifact Evaluation
\end{center}}
]

\appendixAD

\section{Overview of Contributions and Artifacts}

\subsection{Paper's Main Contributions}

\begin{description}
\item[$C_1$] An implementation of a new universal algorithm for one-sided distributed matrix multiplication.
\item[$C_2$] An evaluation of this algorithm on a collection of partitionings, compared to PyTorch DTensor.
\end{description}

\subsection{Computational Artifacts}

\artsampl{
\begin{description}
\item[$A_1$] https://doi.org/10.5281/zenodo.16628051
\end{description}
}

\begin{center}
\begin{tabular}{rll}
\toprule
Artifact ID  &  Contributions &  Related \\
             &  Supported     &  Paper Elements \\
\midrule
$A_1$   &  $C_1$ $C_2$ & Figures 2-3 \\
\bottomrule
\end{tabular}
\end{center}

\section{Artifact Identification}

\newartifact

\artrel

Artifact $A_1$ includes an implementation of the new universal algorithm for one-sided distributed matrix multiplication ($C_1$).  It also includes benchmark code for our algorithm and DTensor that was used to produce the plots in Figure 2 and 3 ($C_2$).

\artexp

Our algorithm achieves faster or comparable performance to DTensor for the matrix sizes shown in Figures 2 and 3.

\arttime

Artifact setup, including installing required software dependencies, is expected to require around 30 minutes.  Artifact execution is expected to require around 2 hours.  Artifact analysis is expected to take around 5-10 minutes.

\artin

\artinpart{Hardware}

Reproducing this artifact requires a system equipped with multiple Intel or Nvidia GPUs as well as an intra-node interconnect such as Xe Link or NVLink.  We used a system equipped with 6 Intel Data Center GPU Max 1550 GPUs connected with Xe Link, similar to the Aurora supercomputer system.

\artinpart{Software}

This artifact depends on Intel SHMEM version 1.4.0, available at \url{https://github.com/oneapi-src/ishmem}.  It also depends on CMake version 3.20 or higher as well as Intel oneAPI 2024.2.0.  On Nvidia systems, CUDA 12.8 and NVSHMEM 3.2.5 are required.  Running the PyTorch DTensor benchmarks requires PyTorch nightly version 2.9.0.dev20250714+xpu or 2.7.1 on Nvidia.

\artinpart{Datasets / Inputs}

No datasets are required.  All matrices are randomly initialized.

\artinpart{Installation and Deployment}

The code for our algorithm should be built with CMake, indicating the correct compiler for the platform.  The \code{DRC_BACKEND} CMake flag should be used to indicate the correct communication backend (Intel SHMEM or NVSHMEM).  On some systems, the \code{ENABLE_MPI} flag may also need to be set to ensure correct launching.  The executable \code{replicated_benchmark} should be built.

Environment variables may need to be set, depending on the system.  We set the following environment variables for Intel platforms:

\begin{center}
\begin{tabular}{l l}
\toprule
  Environment Variable & Value\\
  \midrule
  \code{EnableImplicitScaling} & \code{0}\\
  \code{NEOReadDebugKeys} & \code{1}\\
  \code{FI_PROVIDER} & \code{cxi,tcp;ofi_rxm}\\
  \code{SYCL_DEVICE_FILTER} & \code{:gpu}\\
  \code{FI_CXI_OPTIMIZED_MRS} & \code{0}\\
  \code{FI_CXI_DEFAULT_CQ_SIZE} & \code{131072}\\
  \code{PMI_MAX_KVS_ENTRIES} & \code{10000000}\\
  \code{ISHMEM_SYMMETRIC_SIZE} & \code{8G}\\
  \bottomrule
\end{tabular}
\end{center}

When running DTensor benchmarks, we also use the following environment variables:

\begin{center}
\begin{tabular}{l l}
\toprule
  Environment Variable & Value\\
  \midrule
  \code{CCL_ATL_TRANSPORT} & code{ofi}\\
  \code{CCL_ATL_HMEM} & code{1}\\
  \code{CCL_TOPO_FABRIC_VERTEX_CONNECTION_CHECK} & code{0}\\
  \code{FI_PROVIDER} & code{tcp}\\
  \code{ZE_FLAT_DEVICE_HIERARCHY} & code{FLAT}\\
  \bottomrule
\end{tabular}
\end{center}

On Nvidia systems, we set the following environment variables:

\begin{center}
\begin{tabular}{l l}
\toprule
  Environment Variable & Value\\
  \midrule
  \code{NCCL_ROOT} & Location of NCCL library.\\
  \code{NVSHMEM_HOME} & Location of NVSHMEM library.\\
  \code{NVSHMEM_REMOTE_TRANSPORT} & \code{none}\\
  \code{NVSHMEM_BOOTSTRAP} & \code{MPI}\\
  \bottomrule
\end{tabular}
\end{center}

Different values or additional variables may need to be set depending on system configuration.

The \code{replicated_benchmark} binary should be run using the appropriate launcher for the system.  On Intel platforms, the \code{ishmrun} script must be used.

\artcomp

The first task $T_1$ is to verify that \code{replicated_benchmark} can run successfully.  After this is complete, $T_2$ the benchmark files \code{launch_experiments_mlp1.py} and \code{launch_experiments_mlp2.py} may need to be modified to ensure that they are using the correct launch mechanism for the system.  After this is complete, $T_3$ \code{launch_experiments_mlp1.py} and \code{launch_experiments_mlp2.py} should be run, with their standard output saved to files.  Finally, $T_4$ \code{plot_mlp1.py} and \code{plot_mlp2.py} should be run, taking in the outputs from the MLP-1 and MLP-2 experiments, respectively on standard input.  These plotting scripts will produce the plots shown in Figures 2-3, producing $A_1$.

These tasks are all linearly dependent, that is $T_1 \rightarrow T_2 \rightarrow T_3 \rightarrow T_4$.

If desired, data for additional problem sizes and matrix partitionings can be produced by modifying \code{launch_experiments_mlp1.py} and \code{launch_experiments_mlp2.py}, or by calling \code{replicated_benchmark} directly.

\artout

The plots produced by \code{plot_mlp1.py} and \code{plot_mlp2.py} should match the results observed in $A_1$.  That is, our one-sided algorithm achieves competitive performance compared to DTensor.


\end{document}